 \documentclass[%
preprint,
nofootinbib,
 amsmath,amssymb,
 aps, pra,
]{revtex4-2}

\usepackage{graphicx}
\usepackage{dcolumn}
\usepackage{bm}
\usepackage{bbm}
\usepackage{color}

\usepackage{verbatim}

\usepackage{algorithm}
\usepackage{algpseudocode}

\usepackage[shortlabels]{enumitem}
\usepackage{tensor} 
\usepackage{mathtools}
\newcommand{\bigzero}{\mbox{\normalfont\bfseries 0}}

\begin{document}


\title{Deriving the Landauer Principle From the Quantum Shannon Entropy}

\author{Henrik J. Heelweg}
\email{hheelweg@mit.edu}
 \affiliation{Department of Chemistry, Massachusetts Institute of Technology}
\author{Amro Dodin}%
 \email{adodin@lbl.gov}
\affiliation{%
 Chemical Sciences Division, Lawrence Berkeley National Lab
}%

\author{Adam P. Willard}
 \email{awillard@mit.edu}
\affiliation{Department of Chemistry, Massachusetts Institute of Technology
}%

\date{\today}

\begin{abstract}
We derive an expression for the equilibrium probability distribution of a quantum state in contact with a noisy thermal environment that formally separates contributions from quantum and classical forms of probabilistic uncertainty.
A statistical mechanical interpretation of this probability distribution enables us to derive an expression for the minimum free energy costs for arbitrary (reversible or irreversible) quantum state changes.
Based on this derivation, we demonstrate that - in contrast to classical systems - the free energy required to erase or reset a qubit depends sensitively on both the fidelity of the target state and on the physical properties of the environment, such as the number of quantum bath states, due primarily to the entropic effects of system-bath entanglement.
\end{abstract}

\maketitle
Computing operations, such as the erasure of bits, are implemented as the irreversible transitions of physical systems.
The Landauer principle establishes bounds on the thermodynamics costs of these transitions and information processing more generally.
Landauer argued that the free energy required to erase a single bit of information, expressed in terms of dissipated heat, is given by the inequality, $Q_\mathrm{erase} \geq -T\Delta S_\mathrm{erase}$, where $\Delta S_\mathrm{erase} = k_\mathrm{B} \ln 2$ is the entropy difference between a specific state and a fully randomized state.\cite{Landauer1961}
For a quantum computer, which utilizes qubits rather than classical binary bits, these thermodynamic costs are less easily quantified due to the quantum mechanical nature of qubit states and their ability to entangle with each other and their environments.
In this letter, we present a general formalism for calculating bounds on the thermodynamic costs associated with modifying the states of an open quantum system.
When applied to qubit systems, this formalism yields a quantum Landauer principle.

The primary challenge in formulating a quantum Landauer principle is that quantum system and bath degrees of freedom are generally correlated due to system-bath entanglements. 
Formalizing the contribution of these entanglements to the thermodynamic costs of system state changes is a non-trivial problem.
Previous efforts to solve this problem have derived thermodynamic bounds based on an information-theoretic approach, whereby system-bath unitaries and conservation law arguments are used to express the free energy of qubit state changes as measurable changes in the energy of the bath.\cite{Reeb2014, Esposito2010}
This general approach has the advantage of providing exact solutions that capture the full extent of system-bath correlations.
The drawback of this approach is that finding non-trivial bounds for the dissipated heat that incorporate these correlations requires detailed knowledge of both the qubit transformation and the (typically unknown) initial and final state of the bath. 
In recent years, several studies have been conducted to verify this bound experimentally \cite{Peterson2016, Gaudenzi2018, Toyabe2010, Yan2018}, to make it tighter \cite{Goold2015, Guarnieri2017, Lorenzo2015, Strasberg2017}, or to relax the underlying assumption of initially uncorrelated system-bath states \cite{Jennings2010, Jevtic2015, Micadei2020}.
Our approach, in contrast, does not depend on a complete description of the state of the bath, and has the advantage that it explicitly disentangles the convoluted quantum- and classical-based sources of probabilistic uncertainty to the open quantum system free energy.

The thermodynamic costs of open quantum system state changes include entropic contributions arising from the quantum mechanical nature of the system and potentially its environment. 
One such contribution originates from the intrinsic probabilistic uncertainty of quantum mechanical observables, as exemplified by the phenomenon of wavefunction collapse.
In the context of quantum computing, identically prepared qubit states can yield different outcomes upon observation.
Another contribution originates from the ability of quantum systems to entangle with their surroundings.
The information content of an entangled qubit state can thus be obscured - encoded in the properties of external quantum correlations.

In our formalism, we will formally distinguish these quantum forms of uncertainty from the (more traditional) classical forms of uncertainty.
We illustrate this distinction by considering an idealized quantum circuit that when in isolation is capable of producing the exact same quantum state every time it is run.
Even in isolation, uncertainty remains in the measured outcome of this circuit due to the probabilistic nature of wavefunction collapse.
For the isolated circuit, the only source of statistical uncertainty in the measured outcome is due to the quantum mechanical nature of the circuit.
We refer to this source of uncertainty as the \textit{intrinsic quantum uncertainty}.
This intrinsic quantum uncertainty is not limited to isolated systems, as the observable properties of composite system-bath wave functions are subject to an analogous effect.

On the other hand, classical uncertainty is a consequence of ensemble statistics, \textit{e.g.}, due to thermal fluctuations of the environment.
In the case where the perfect quantum circuit interacts with uncontrolled environmental degrees of freedom, every time the circuit runs, external noise may cause it to generate a different quantum state.
The distribution of these quantum states represents a different - purely classical - source of statistical uncertainty in the measured outcome of the circuit.
We refer to this ensemble-based source of uncertainty as the \textit{classical uncertainty}.
There is no classical uncertainty for the idealized isolated quantum circuit.

Upon measurement, the overall uncertainty of an ensemble quantum system is a convolution of these classical and quantum sources.
The convoluted statistics of the measured outcomes are most commonly represented in terms of an ensemble density matrix, $\rho$.
Alternately, the quantum and classical sources of statistical uncertainty can be represented separately via a classical probability distribution of quantum density matrices, each representing a possible outcome of an individual quantum system.\cite{Dodin2019, Dodin2021, Dodin2022, Carollo2019, Carollo2021, Anderson2022, Anderson2024, Schile2018}
In this work, we adopt this alternate approach to representing observational uncertainty, hereby referred to as the $P$-ensemble approach.

We start by deriving the equilibrium $P$-ensemble distribution for a quantum system in contact with a noisy thermal environment.
Every unique configuration of this system-bath composite can be described by a single-configuration density matrix, $\Gamma$, that jointly specifies all circuit and bath degrees of freedom.
We define the probability for any specific configuration to arise via the classical probability distribution, $P[\Gamma]$.
The standard ensemble density matrix, $\rho$, is related to $P[\Gamma]$ via the expression,
\begin{align}
\rho = \int [d\Gamma] \Gamma P[\Gamma]\equiv \langle \Gamma \rangle_{P},
\label{eq:RhoPensembleRelation}
\end{align}
where $\int [d\Gamma]$ denotes the integration with respect to the uniform measure on the state space of composite density matrices, and $\langle \cdot\rangle_P$ denotes the distribution average over $P[\Gamma]$.

Many different distributions, $P[\Gamma]$, can map to the same ensemble density matrix $\rho$.
We determine the correct distribution based on a Jayne's-type maximum entropy argument.\cite{Jaynes1957}
Specifically, the \textit{quantum Shannon entropy} of a given distribution is,
\begin{align}
\mathcal{S} \equiv -\int [d\Gamma] P[\Gamma]\ln P[\Gamma],
\label{eq:quantumShannon}
\end{align}
which can be directly interpreted in terms of the thermodynamic entropy, as we demonstrate in section B of the Appendix.
We assert that the physically correct $P$-ensemble distribution is that which maximizes $\mathcal{S}$ subject to the constraint that the equilibrium ensemble density matrix corresponds to the Gibbs state,\cite{Jaynes1968} \textit{i.e.},  
\begin{align}
\rho = \rho_\mathrm{G} \equiv \frac{e^{-\beta \hat{H}}}{\mathrm{Tr}\{e^{-\beta \hat{H}}\}},
\label{eq:GibbsConstraint}
\end{align}
where $\hat{H}$ is the composite system (circuit and bath) Hamiltonian, and $\beta=(k_\mathrm{B} T)^{-1}$, where $k_\mathrm{B}$ is the Boltzmann constant, and $T$ the temperature.
This constraint ensures that the eigenstates of $\hat{H}$ appear with their appropriate Boltzmann weights.

The distribution $P[\Gamma]$ that maximizes $\mathcal{S}$ subject to the Gibbs state constraint, as derived in section A of the Appendix, is a Boltzmann-like distribution,
\begin{align}
P[\Gamma]=\frac{e^{-\beta'\bar{H}[\Gamma]}}{\int [d\Gamma] e^{-\beta'\bar{H}[\Gamma]}}\equiv \frac{e^{-\beta'\bar{H}[\Gamma]}}{Z}
\label{eq:EquilibriumDistribution}
\end{align}
where $\bar{H}[\Gamma] \equiv \mathrm{Tr}\{\hat{H}\Gamma\}$ is the quantum-average energy of state $\Gamma$, $Z$ is the $P$-ensemble partition function, $\beta'=(k_\mathrm{B}\Theta)^{-1}$ and $\Theta$ is the statistical temperature. 
 We assign the value of $\Theta$ to satisfy the constraint of Park and Band,\cite{Park1977} that both $\rho$ and $P[\Gamma]$ yield consistent mean ensemble energies, \textit{i.e.}, that $\bar{H}[\rho_\mathrm{G}]=\langle \bar{H}[\Gamma]\rangle_P$.

We now leverage the $P$-ensemble description of the system-bath composite to derive a formalism for expressing the equilibrium statistics of the reduced system density matrix.
We assume that the system and bath occupy separate Hilbert spaces, $\mathcal{H}_\mathrm{S}$ and $\mathcal{H}_\mathrm{B}$, respectively.
Each composite microstate can be represented by a pure state density matrix, $\Gamma \in \mathcal{H}=\mathcal{H}_\mathrm{S} \otimes \mathcal{H}_\mathrm{B}$.
For a given composite microstate, $\Gamma$, the system state can be expressed in its own Hilbert space ($\mathcal{H}_\mathrm{S}$) via the reduced density matrix, by tracing out the bath, \textit{i.e.}, $\sigma=\mathrm{Tr_B}\{\Gamma\}$.
Thus, each composite microstate, $\Gamma$, maps to a specific reduced circuit density matrix, $\sigma$, but multiple different composite states may map to the same $\sigma$.
We define the set of all composite microstates that map to a given $\sigma$ as $C_\sigma \equiv \{\Gamma \in \mathcal{H} : \mathrm{Tr_B}\{\Gamma\}=\sigma\}$.

With this formalism, we can develop a statistical mechanical interpretation of quantum circuit states.
The probability to observe a given circuit state is,
\begin{equation}
\label{eq:Pofsigma}
    P[\sigma] = \frac{1}{Z} \int_{C_\sigma} [d\Gamma] e^{-\beta' \bar{H}[\Gamma]} = \frac{Z[\sigma]}{Z} \equiv \frac{1}{Z}e^{-\beta' F[\sigma]},
\end{equation}
where the integral extends over the region of $\mathcal{H}$ defined by the set of microstates $C_\sigma$, and we have defined $F[\sigma]$ as the free energy of a given circuit state.
We aim to express this free energy as a sum of energetic and entropic components, \textit{i.e.}, $F[\sigma] = E[\sigma] - TS[\sigma]$.
To accomplish this, we decompose the composite Hamiltonian into a general system-bath form, $\hat{H} = \hat{H}_\mathrm{S} + \hat{H}_\mathrm{B} + \hat{V}$, where $\hat{H}_\mathrm{S} \in \mathcal{H}_\mathrm{S}$, and $\hat{H}_\mathrm{B} \in \mathcal{H}_\mathrm{B}$ are the circuit and bath Hamiltonians, respectively, and $\hat{V}\in \mathcal{H}$ describes the circuit-bath coupling.
With this decomposition, we can isolate the mean field energetic contributions of the bath to state $\sigma$ as,
\begin{equation}
\label{eq:EnthalpyEntropyA}
E[\sigma] \equiv \int_{C_\sigma} [d\Gamma ] \bar{H}[\Gamma] P[\Gamma \vert \sigma] = \bar{H}_\mathrm{S}[\sigma] + \langle \bar{H}_\mathrm{B}\rangle_\sigma + \langle \bar{V}\rangle_\sigma, 
\end{equation}
where $P[\Gamma \vert \sigma] = e^{-\beta'\bar{H}[\Gamma]}/Z[\sigma]$ is the conditional probability of state $\Gamma$ within the set of states defined by $C_\sigma$, $\langle \cdot \rangle_\sigma$ is the corresponding conditional average, and where we have made use of the decompositions $\bar{H}_\mathrm{B}[\Gamma] = \langle \bar{H}_\mathrm{B} \rangle_\sigma + (\Delta \bar{H}_\mathrm{B}[\Gamma])_\sigma$ and $\bar{V}[\Gamma] = \langle \bar{V} \rangle_\sigma + (\Delta \bar{V}[\Gamma])_\sigma$ (see section C of the Appendix).
The associated entropic contribution to the free energy is thus given by
\begin{equation}
\begin{aligned}
        \label{eq:EnthalpyEntropyB}
    TS[\sigma] &\equiv - k_\mathrm{B}\Theta \int_{C_\sigma}[d\Gamma] P[\Gamma \vert \sigma] \ln P[\Gamma \vert \sigma] \\
    &= k_\mathrm{B} \Theta \ln \left( \int_{C_\sigma}[d\Gamma]e^{-\beta'((\Delta \bar{H}_\mathrm{B}[\Gamma])_{\sigma}
        +(\Delta \bar{V}[\Gamma])_{\sigma})} \right). 
\end{aligned}
\end{equation}
Alternatively, the system free energy can also be decomposed into internal and external contributions,
\begin{equation}
    F[\sigma] = \bar{H}_\mathrm{S}[\sigma] + \mathcal{F}_\mathrm{solv}[\sigma],
\end{equation}
where $\mathcal{F}_\mathrm{solv}[\sigma]$ defines the free energy to incorporate (solvate) the system within the bath.

\noindent
Equations~\ref{eq:Pofsigma},~\ref{eq:EnthalpyEntropyA} and~\ref{eq:EnthalpyEntropyB} establish a formalism for quantifying the free energy cost for changing the state of a quantum system in contact with a bath.
Even in cases where $\bar{H}_\mathrm{S}[\sigma]$ is trivial (such as the spin-boson model), the remaining terms in the solvation free energy, $\mathcal{F}_\mathrm{solv}[\sigma]$, can be challenging to calculate.
Still, analytical solutions can be derived in the regime where the bath is weakly coupled to the system, \textit{i.e.}, $\sigma_\mathrm{G}=\mathrm{Tr_B}\lbrace \rho_\mathrm{G} \rbrace$ is negligibly influenced by the presence of the bath, and the number of quantum degrees of freedom of the bath, $M\equiv \mathrm{dim}(\mathcal{H}_\mathrm{B})$, is much larger than that of the system, $N\equiv \mathrm{dim}(\mathcal{H}_\mathrm{S})$.

\begin{figure*}[b]
    \centering
    \includegraphics[width=.9\textwidth]{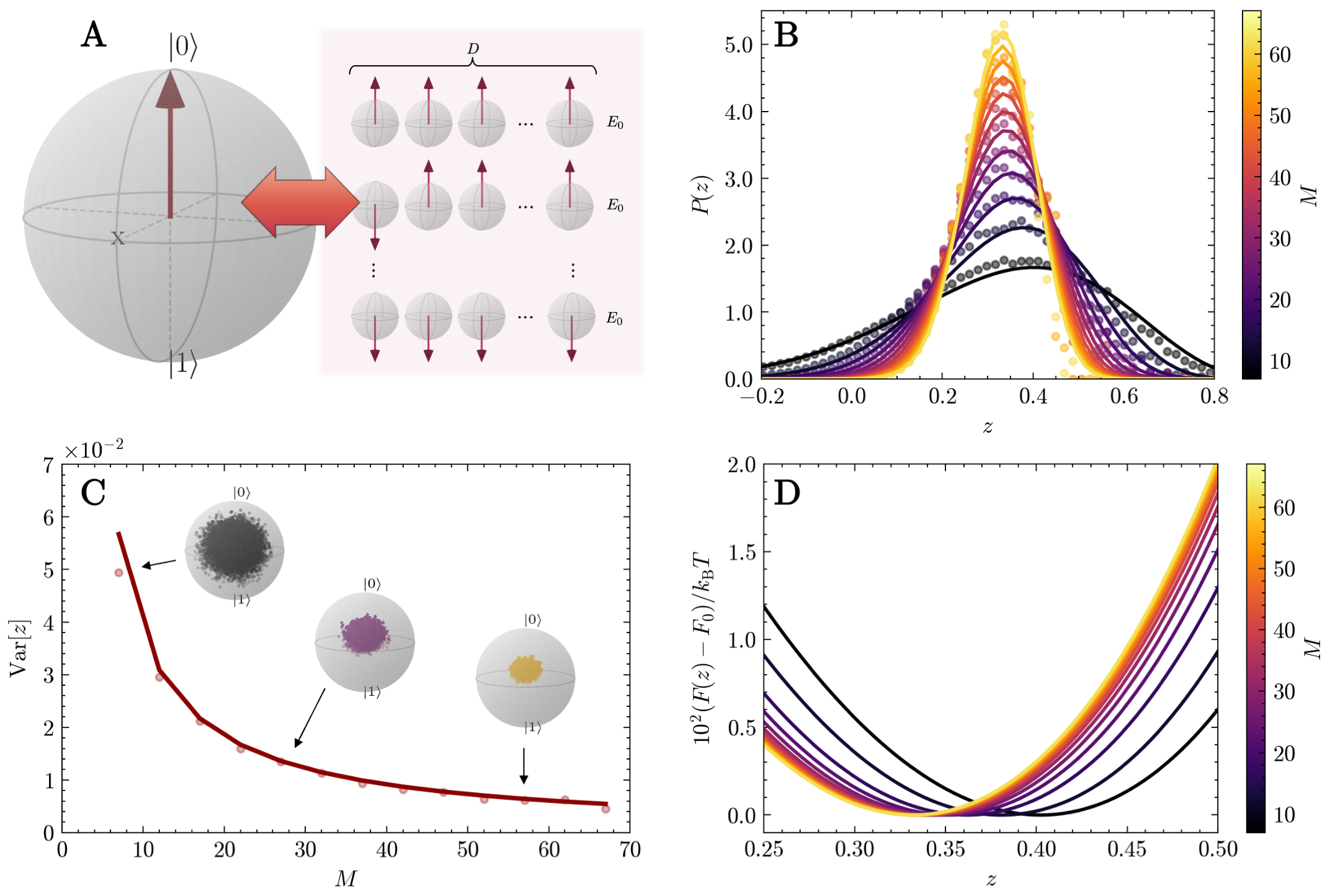}
    \caption{
    A qubit circuit weakly coupled to $D$ energetically unbiased bath qubits.
    (A) An illustration of the system-bath composite, where the system qubit state is represented on a Bloch sphere and the $M=2^D$ bath states each have identical energies, $E_0$.
    (B) The equilibrium probability distribution for the system qubit to have a Bloch sphere coordinate with value $z$, plotted for varying bath sizes.
    Points indicate numerical results and lines indicate the analytical solution based on Eq.~\ref{eq:degenerateFreeEnergy}.
    (C) Variance of the distribution $P(z)$ as a function of $M$ for numerical (points) and analytical (line) results.
    Insets show an equilibrium sample of the system state (each represented as a point in the Bloch sphere).
    (D) $F(z)=-k_\mathrm{B}T \ln P(z)$ for analytical solutions where $F_0\equiv \min F(z)$. } 
    \label{fig:degenerateBath}
\end{figure*}

We first consider the case of an energetically degenerate bath.
In a model with $M$ degenerate bath states, the solvation free energy for $\sigma$ is purely entropic, yielding a system free energy of,
\begin{equation}
\label{eq:degenerateFreeEnergy}
\begin{aligned}
    F[\sigma]= & \bar{H}_\mathrm{S}[\sigma]-k_\mathrm{B}\Theta \Bigl[ \ln \mathrm{det}^{M-N}[\sigma] + \ln \left(2^{-N} \mathrm{Vol} \left(\mathcal{V}_N(\mathbb{C}^M)\right) \right) \Bigl],
\end{aligned}
\end{equation}
where $\mathcal{V}_N(\mathbb{C}^M) \equiv  \{ \tilde{X} \in \mathbb{C}^{M \times N} : \tilde{X}^\dagger \tilde{X}=\mathbbm{1}_N\}$ is the complex Stiefel manifold, a topological manifold formed by $M\times N$ semi-unitary matrices, and $\mathrm{Vol} \left(\mathcal{V}_N(\mathbb{C}^M)\right) $ its associated volume over the Haar measure of that group (see section D1 of the Appendix). 
The volume of the Stiefel manifold reflects the geometry of the underlying subset $C_\sigma$ on the composite Hilbert space.
The determinant of a qubit is directly related to its purity, $\mathrm{Pur}[\sigma] \equiv \mathrm{Tr}\{\sigma^2\}$, by $\mathrm{det}[\sigma] = (1-\mathrm{Pur}[\sigma])/2$. 
A reduction in purity is therefore associated with a lowering of the free energy.
In the limiting case when $\sigma$ is a pure state, $F[\sigma]$ diverges to infinity. 
This finding represents the thermodynamic counterpart of an argument by Reeb and Wolf about the bound on the pureness of quantum states in a thermal reservoir.\cite{Reeb2014}   

To put these analytical results in context, we consider the numerical simulation a composite model consisting of a single qubit (a primitive quantum circuit) in contact with a degenerate bath of $D$ qubits.
In this model, the circuit is governed by the Hamiltonian $\hat{H}_\mathrm{S}=-\frac{\hbar \omega_0}{2}\hat{\sigma}_z$, represented in the $\{\vert 0 \rangle, \vert 1 \rangle\}$ logical basis, and thus biased toward the $\vert 0 \rangle$ state.
The bath qubits are unbiased so that all bath states have identical potential energy $E_0$, as illustrated in Fig.~\ref{fig:degenerateBath}A. 
We use this model to study the effect of the bath size, $M=2^D$, on the statistics of the circuit state, quantified in terms of the observable $z \equiv\bar{\sigma_z}[\sigma]=\mathrm{Tr}\{\sigma_z \sigma\}$.
The numerical results are derived from Markov Chain Monte Carlo (MCMC) simulations as described in section E of the Appendix.

The results, presented in Fig.~\ref{fig:degenerateBath}B, are in excellent agreement with the analytical solution to $F[\sigma]$.
When $M$ is small, the distribution of circuit qubit states is broadly distributed along the $z$-axis with a asymmetry that reflects the energetic bias for the circuit qubit toward the $\sigma = \vert 0 \rangle \langle 0 \vert$ state.
As $M$ increases, the distribution of circuit states narrows, reflecting the increasing influence of the system-bath solvation entropy on the circuit state statistics.
More specifically, system-bath entanglement reduces the purity of the system state.
Pure system states reside on the surface of the Bloch sphere, while entangled states reside within the Bloch sphere, with reduced Bloch vector magnitude and thus reduced magnitude of $z$.
This effect is illustrated in the insets of Fig.~\ref{fig:degenerateBath}C.

Following Landauer's insight, we thus leverage our formalism to specify a bound on the heat dissipation associated with the erasure or preparation of a quantum circuit.
Equation~\ref{eq:degenerateFreeEnergy} can be utilized to compute the entropy change in transforming the circuit from state $\sigma_\mathrm{i}$ to state $\sigma_\mathrm{f}$ in the presence of a noisy, degenerate, and weakly-coupled environment.
Based on our results, we find for any such operation, the heat dissipation into the degenerate spin bath, $\Delta Q_\mathrm{B} \geq -T\Delta S$, has a lower bound given by the Second Law, 
\begin{align}
    \Delta Q_\mathrm{B}/k_\mathrm{B} \geq \Theta\ln \Bigl( \frac{\det[\sigma_\mathrm{i}]}{\det[\sigma_\mathrm{f}]}\Bigr)^{M-N}.
    \label{eq:degenerateLandauerBound}
\end{align}
The underlying thermodynamic argument allows us to interpret Eq.~\ref{eq:degenerateLandauerBound} as an equality for any unitary operation on the circuit that is conducted in a physically reversible manner. \par

\begin{figure}[t]
    \centering
    \includegraphics[width=.45\textwidth]{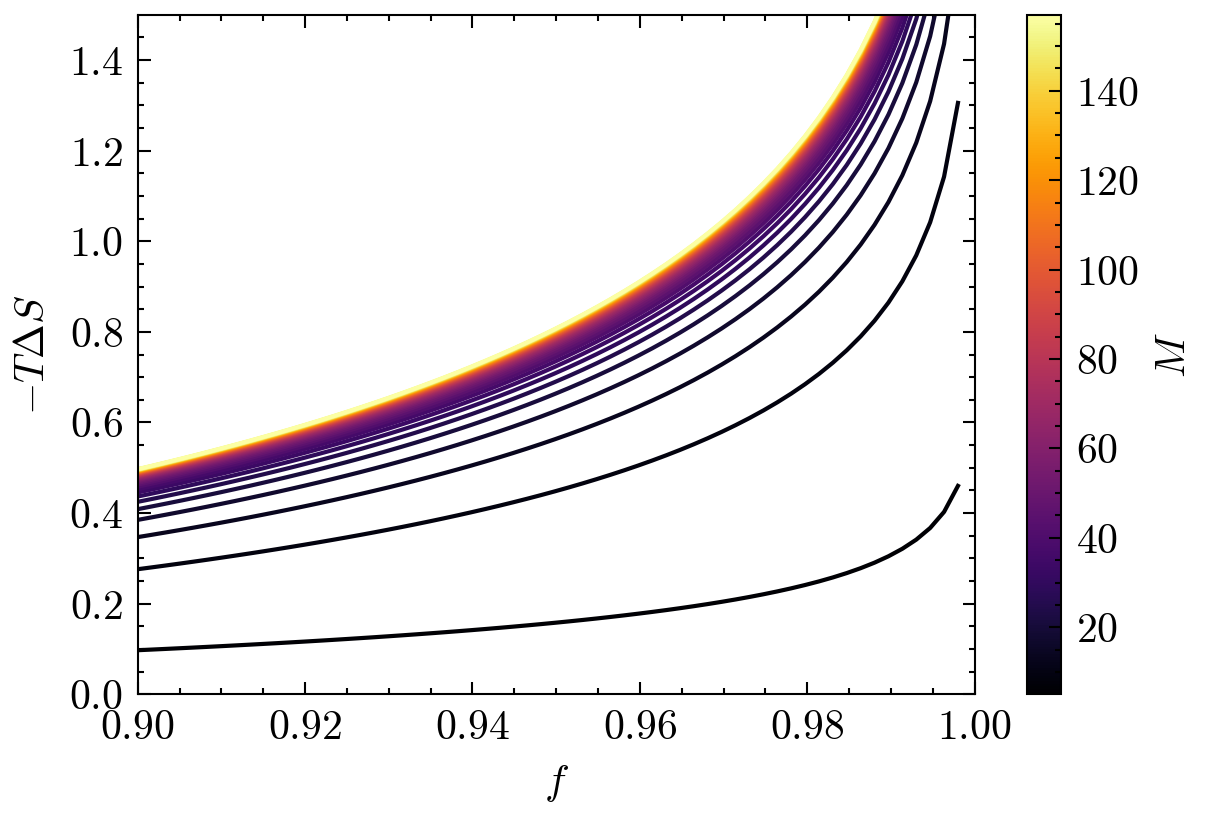}
    \caption{Landauer bound Eq.\ \eqref{eq:degenerateLandauerBound} for erasing a two-qubit Bell state with increasing fidelity $f$ for various dimensions of a (weakly) coupled $M$-degenerate bath.} 
    \label{fig:LandauerTwoQubits}
\end{figure}

To highlight the implications of Eq.~\ref{eq:degenerateLandauerBound}, we consider a system of two qubits ($N=4$) that is initially in a fully equilibrated state (maximally entangled with the bath), \textit{i.e.}, $\sigma_\mathrm{i}=\mathrm{diag}(1/N, ..., 1/N)$, that is restored to a lower entropy entangled Bell state with fidelity $f$, \textit{i.e.}, $\sigma_\mathrm{f}(f)= 2(1-f)\sigma_\mathrm{i}+(2f-1)\sigma_\mathrm{B}$ with $\sigma_\mathrm{B}=\mathrm{diag}(1/2, 0, 0, 1/2)$.
The entropy change, and thus the minimal required heat dissipation, is plotted in Fig.~\ref{fig:LandauerTwoQubits}.
We find that achieving a higher fidelity final state requires an increased exchange of heat with the environment, and that the amount of heat exchange depends sensitively on the size of the bath.
The entropic barrier associated with eliminating system-bath entanglement (\textit{i.e.}, the effective desolvation of the two-qubit system) grows rapidly with the size of the bath.
This bath size dependence represents a stark contrast to the classical Landauer principle, which is bath-independent.

We now consider a non-degenerate bath model to evaluate the influence of bath energetic fluctuations on the circuit solvation free energy.
If $\Lambda(\hat{H}_\mathrm{B}) = \{\lambda_i\}_{i=1}^M$ is the spectrum of $\hat{H}_\mathrm{B}$, with $\lambda_1 < ... < \lambda_M$, and $\Lambda(\sigma)=\{\eta_i\}_{i=1}^N$ the spectrum of the circuit state, with $\eta_1 < ... < \eta_N$, then the free energy of any circuit state, $\sigma$, is given by (see section D2 of the Appendix),
\begin{equation}
\label{eq:thermalFreeEnergy}
\begin{aligned}
    F[\sigma]=
    \bar{H}_\mathrm{S}[\sigma]-k_\mathrm{B}\Theta \Biggl[  \ln \left( \frac{\vert \mathrm{det} [I(\sigma)]\vert}{\Delta(\sigma)}  \right)
    +\ln \left( \frac{2^{-N} \mathrm{Vol} \left(\mathrm{V}_N(\mathbb{C}^M)\right)\prod_{i=M-N}^M i!}{\Delta(H_\mathrm{B})} \right)\Biggr],
\end{aligned}
\end{equation}
with the $M\times M$-dimensional matrix,
\begin{equation}
\begin{aligned}
    I(\sigma) \equiv [1, (\beta'\lambda_i), ...,  (\beta'\lambda_i)^{M-N-1}, e^{-\beta'\eta_1 \lambda_i},
    ..., e^{-\beta'\eta_N \lambda_i}]_{i=1,...,M},
\label{eq:interactionMatrix}
\end{aligned}
\end{equation}
and with ,
\begin{subequations}
\label{eq:vandermondeDeterminants}
\begin{align}
    \Delta(\sigma) &\equiv \prod_{1 \leq i < j \leq N}\left( \eta_j-\eta_i \right) \\
     \Delta(H_\mathrm{B}) &\equiv \prod_{1 \leq i < j \leq M}\left( \beta'\lambda_j-\beta'\lambda_i \right).
\end{align}
\end{subequations}
While strictly only being valid for a non-degenerate bath spectrum $\Lambda(\hat{H}_\mathrm{B})$, Eq.~\ref{eq:thermalFreeEnergy} can be extended to account for bath degeneracies by repetitive application of L'Hôpital's rule. 

\begin{figure*}[ht]
    \centering
    \includegraphics[width=.9\textwidth]{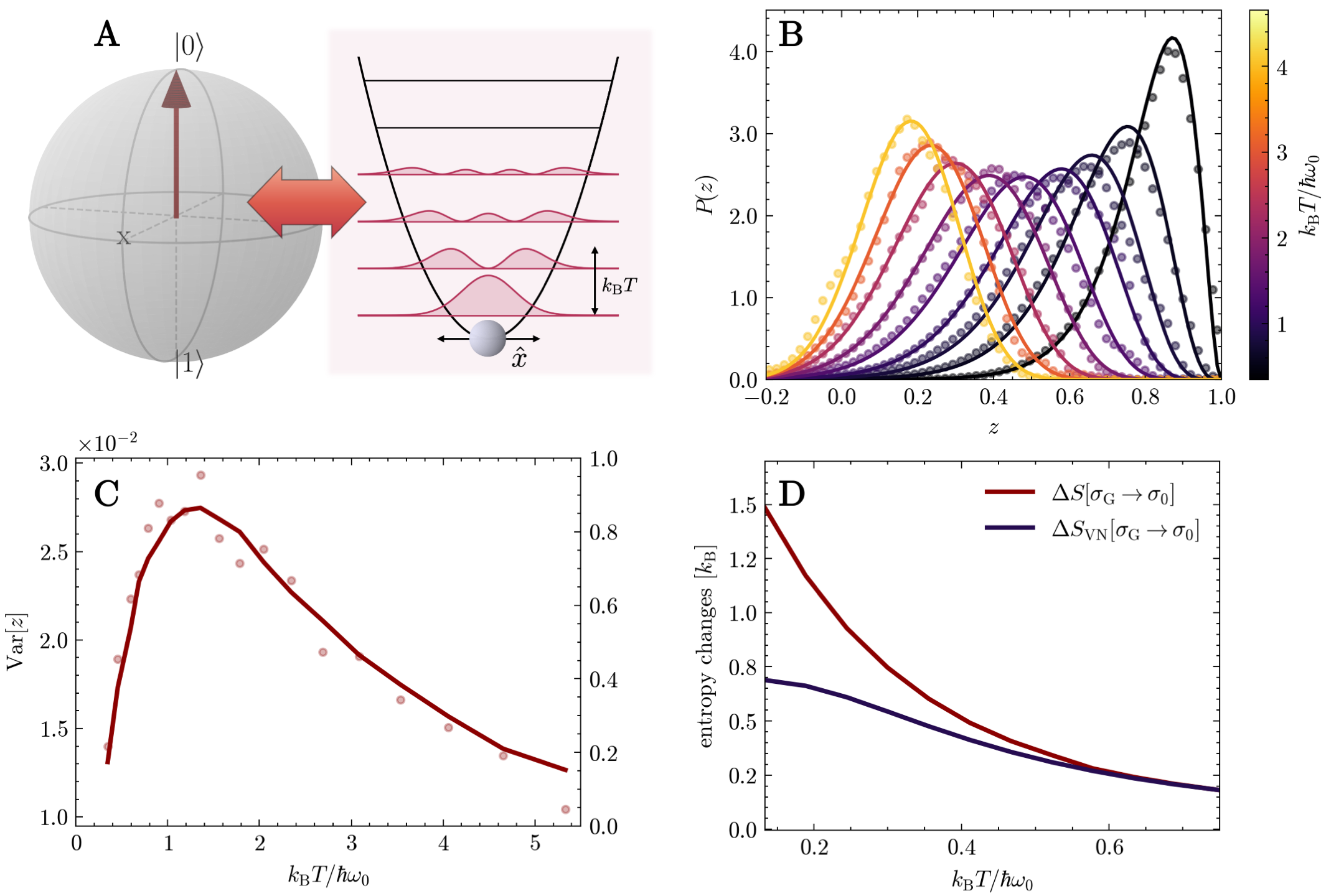}
    \caption{
    A qubit circuit weakly coupled to a truncated harmonic oscillator with $\hbar\omega=0.2 \hbar\omega_0$.
    (A) An illustration of the circuit-bath composite.
    (B) The equilibrium probability distribution for the system qubit to have a Bloch sphere coordinate with value $z$, plotted for different temperatures.
    Points indicate numerical results and lines indicate the analytical solution based on Eq.~\ref{eq:thermalFreeEnergy}.
    (C) Variances of the obtained analytical (lines) and numerical (points) distributions with increasing temperature. 
    (D) Landauer bound for the erasure of a maximally entangled qubit in a thermalized environment as obtained from Eq.~\ref{eq:thermalLandauerBound}, in comparison with the von Neumann entropy, $\Delta S_\mathrm{VN}$. 
    } 
    \label{fig:thermalBath}
\end{figure*}

We again use MCMC simulations (see section E of the Appendix) to sample the equilibrium states of a qubit circuit, governed by the Hamiltonian $\hat{H}_\mathrm{S}=-\frac{\hbar \omega_0}{2}\hat{\sigma}_z$, but this time coupled to the displacement of a single thermally truncated harmonic oscillator with $M$ distinct energy levels (Fig.\ \ref{fig:thermalBath}A and B).  
Once again, we find excellent agreement between the numerical and analytical results.
At low temperatures ($\hbar \omega_0 > k_\mathrm{B} T$), only a few oscillator levels are thermally populated and the equilibrium statistics of the circuit are largely a result of the enthalpic driving force towards lower-energy states, driven primarily by the properties of $\hat{H}_\mathrm{S}$, due to the weak system-bath coupling.
As temperature increases and more oscillator states become thermally accessible, the distribution $P(z)$ is modified by two competing effects:
(1) The distribution broadens and shifts to lower values of $z$ due to entropic and enthalpic forces, respectively, from the qubit system, and 
(2) the distribution narrows as the entropic contributions of system-bath entanglement increasingly manifest.
Due to the competition of these two effects, the distributional variance of the qubit is non-monotonic, peaking at a specific temperature (Fig.\ \ref{fig:thermalBath}C).
The width of the distribution increases initially due to classical entropic effects of the qubit and narrows again due to the effects of entanglement.
This non-monotonic trend in the variance of the system distribution with temperature is a purely quantum-mechanical effect.

Using the analytical results, we can quantify the thermodynamic bounds on the quantum state change, $\sigma_\mathrm{i} \rightarrow \sigma_\mathrm{f}$, this time including a weakly coupled non-degenerate bath, as, 
\begin{equation}
\begin{aligned}
    \Delta Q_\mathrm{B}/k_\mathrm{B} & \geq  \Theta\ln \Bigl( \frac{\vert \det[I(\sigma_\mathrm{i})]\vert \Delta(\sigma_\mathrm{f})}{\vert \det[I(\sigma_\mathrm{f})]\vert \Delta(\sigma_\mathrm{i})}\Bigr) \\
    &-\frac{1}{k_\mathrm{B}} \Bigl[ \mathrm{Tr} \bigl\{ I(\sigma_\mathrm{i})^{-1}I(\sigma_\mathrm{i})'\bigr\}-\mathrm{Tr} \bigl\{ I(\sigma_f)^{-1}I(\sigma_f)'\bigr\}\Bigr],
    \label{eq:thermalLandauerBound}
\end{aligned}
\end{equation}
where $I(\sigma)' \equiv \partial I(\sigma)/\partial \beta'$ is a scalar derivative of matrix $I(\sigma)$, and $I(\sigma)^{-1}$ its inverse (see section D3 of the Appendix). 
We can use this bound to estimate the minimally dissipated heat for restoring a maximally entangled qubit state, $\sigma_i$, to a pure state, e.g. setting an ancilla qubit in the quantum memory back to $\vert 0 \rangle$ at the end of an algorithm. 
We here assume that this ancilla qubit has been fully thermalized through prolonged contact with the surrounding to reach the Gibbs state, \textit{i.e.}, $\sigma_f = \sigma_\mathrm{G}$.
The entropy change for such a bit erasure is plotted in Fig.\ \ref{fig:thermalBath}D as a function of temperature.

At high temperatures with a large number of thermally accessible bath states, the bound for the dissipated heat coincides with the von Neumann entropy, $\Delta S_\mathrm{VN}$, which corresponds to the classical Landauer bound. 
This result has been previously demonstrated both experimentally,\cite{Yan2018} and theoretically,\cite{Reeb2014} and aligns with the well-established notion that quantum effects disappear in the high-temperature regime, \textit{i.e.}, the qubit behaves like a classical (discrete) bit. 
In the low-temperature regime, there is a large entropic bottleneck that must be overcome to eliminate system-bath entanglement and generate an unentangled pure state.

In summary, we have developed a thermodynamic framework for quantifying bounds on the heat that a quantum system undergoing an arbitrary (reversible or irreversible) transformation exchanges with its environment.
This framework - essentially a quantum Landauer principle - can be utilized to evaluate and optimize the efficiencies of quantum technologies and suggest design rules for quantum system engineering.
It may be possible, through a type of calorimetry, to leverage this framework to infer the difficult-to-measure details of quantum system-bath interactions.
We formally present a new approach to quantum information theory that casts quantum statistics in the same mathematical language as classical statistical mechanics.
In doing so, the $P$-ensemble method provides new foundational insights into quantum information and the distinction between quantum and classical systems.

\newpage
\appendix

\section{$P$-Ensemble Equilibrium Distribution} \label{sec:PEnsembleEquil}
According to the usual formulation of thermodynamic constraints in open quantum system as obtained by a entropy maximization argument performed on the $\rho$-ensemble, the ensemble averaged density matrix is given by the Gibbs state
\begin{align}
    \rho_\mathrm{G} = \int [d\Gamma] \Gamma P[\Gamma] = \langle \Gamma \rangle_P = \frac{\exp(-\beta \hat{H})}{\mathrm{Tr}\{\exp(-\beta \hat{H})\}}.
    \label{eq:GibbsState}
\end{align}
Eq.~\ref{eq:GibbsState} ensures that observables over the whole ensemble, specified by a underlying distribution $P[\Gamma]$ over the space of single-system density matrices, are consistent with thermodynamics, with each energy eigenstate appearing with the appropriate Boltzmann weight in equilibrium. 
In a subjective interpretation of probabilities, $P[\Gamma]$ reflects the human ignorance as to the true state of the quantum system if one has been given an ensemble.
In principle, there are infinitely many distributions $P[\Gamma]$ that are consistent with the constraint in Eq.~\ref{eq:GibbsState}: 
One may, for example, construct a $\delta$-function centered on the Gibbs state, $\rho_\mathrm{G}$, indicating that every individual system is identically prepared in that state. 
Notably however for the quantum domain, even in that limit of full knowledge about the true system state, measurement of such a state will still lead to probabilistic measurement outcomes due to the inherent quantum uncertainty encoded in $\Gamma$.

\subsection{Entropy Maximization Subject to Constraints}
In the spirit of classical statistical mechanics, we want to consider the distribution for $P[\Gamma]$ that assumes the least about the underlying statistics, \textit{i.e.}, the maximum entropy distribution consistent with the Gibbs condition (Eq.~\ref{eq:GibbsState}).
The entropy as an information measure for $P[\Gamma]$ on the phase space of mutually exclusive state representations $\Gamma$, $\mathcal{S}$, has been defined in the main text as an integral $\int[d\Gamma]$ over all single-system density matrices for a system-bath composite, where $[d\Gamma]$ is the invariant measure on that space. 
If we denote with $\Omega \equiv \{\Gamma \in \mathbb{C}^{p \times p}: \Gamma = \Gamma^\dagger, \Gamma >0, \mathrm{Tr}\{\Gamma\}=\mathrm{Tr}\{\Gamma^2\}=1\}$ the phase space of (pure) system-bath composite states $\Gamma$, then $\mathcal{S}$ can also be written as a discrete sum over an infinitely dense grid $\mathcal{G}$ on $\Omega$, \textit{i.e.},
\begin{align}
    \mathcal{S} \equiv -\int[d\Gamma] P[\Gamma] \ln P[\Gamma] = -\sum_{\Gamma \in \mathcal{G}} P[\Gamma] \ln P[\Gamma].
\end{align}
Such a discretization of phase space allows us to perform the entropy maximization on an arbitrary set of points on $\mathcal{G} \subset \Omega$ in a discretized fashion, imposing the following properties:
\begin{enumerate}
    \item {
    \textbf{Closure under Real Reflection:} For all $\Gamma \in \mathcal{G}$, the real reflected state $\Gamma^{(ij, R)} \in \mathcal{G}$, where $\Gamma^{(ij,R)}$ is obtained by inverting the sign of $\Gamma_{ij}^R$, \textit{i.e.}, $\Gamma_{ij}^R \rightarrow -\Gamma_{ij}^R$. 
    }
    \item {
    \textbf{Closure under Imaginary Reflection:} For all $\Gamma \in \mathcal{G}$, the real reflected state $\Gamma^{(ij, I)} \in \mathcal{G}$, where $\Gamma^{(ij,I)}$ is obtained by inverting the sign of $\Gamma_{ij}^I$, \textit{i.e.}, $\Gamma_{ij}^I \rightarrow -\Gamma_{ij}^I$.
    }
    \item {
    \textbf{Closure under Population Transfer:} For all $\Gamma \in \mathcal{G}$ the population transposed state $\Gamma^{(ij)} \in \mathcal{G}$, where $\Gamma^{(ij)}$ is obtained by transposing the population of states $i$ and $j$, \textit{i.e.}, $\Gamma_{ii} \leftrightarrow \Gamma_{jj}$.
    }
\end{enumerate}

\paragraph{Incorporating the Gibbs Constraint.}
Due to the rotational invariance of the measure $[d\Gamma]$, we can select an arbitrary basis for the matrix representation of $\rho_\mathrm{G}$ and $\Gamma$ in Eq.~\ref{eq:GibbsState}.
It is convenient to select as a basis $\{\vert i \rangle\}_{i=1}^p$ the energy eigenbasis of the Hamiltonian $\hat{H}$ since $\rho_\mathrm{G}$ becomes diagonal. 
For each of the $p-1$ population and $p(p-1)$ real coherence degrees of freedoms of the Hermitian density matrix $\rho_\mathrm{G}$, we can define separate constraint. 
Denoting the population components of $\Gamma$ as $\Gamma_{ii} \equiv \langle i \vert \Gamma \vert i \rangle$, and the real and imaginary parts of the coherence by $\Gamma_{ij}^R \equiv \Re [\langle i \vert \Gamma \vert j \rangle]$ and $\Gamma_{ij}^I \equiv \Im [\langle i \vert \Gamma \vert j \rangle]$, respectively, Eq.~\ref{eq:GibbsState} can be re-expressed as
\begin{subequations}
    \begin{align}
        \langle \Gamma_{ij}^R \rangle_P =  \langle \Gamma_{ij}^I \rangle_P = 0 \label{eq:coherenceConstraints}\\
        \langle \Gamma_{ii} \rangle_P = \frac{e^{-\beta E_i}}{\sum_{i=1}^p e^{-\beta E_i}}. \label{eq:populationConstraints}
    \end{align}
    \label{eq:matrixConstraints}
\end{subequations}
We denote with $\eta_{ij}^R$ and $\eta_{ij}^I$ the Lagrange multipliers responsible for maintaining the (real and imaginary) coherence constraints (see Eq.~\ref{eq:coherenceConstraints}), and with $\mu_i$ the Lagrange multipliers responsible for satisfying the population constraints (see Eq.~\ref{eq:populationConstraints}).
In order to ensure that the discretized set of probabilities, $\{P[\Gamma]\} \equiv \{P[\Gamma] : \Gamma \in \mathcal{G}\}$, is normalized, \textit{i.e.}, $\sum_{\Gamma \in \mathcal{G}}P[\Gamma] = 1$, we introduce the Lagrange multiplier $\lambda_0$. 

\paragraph{Constrained Entropy Maximization.}
With those constraints, the entropy maximization problem can be written as 
\begin{equation}
\begin{aligned}
    \{P[\Gamma]\} =\underset{\{P[\Gamma]\}}{\arg\!\max} \Biggl( &
    - \langle \ln P[\Gamma] \rangle_P
    -\sum_{i < j}\left(\eta_{ij}^R \langle \Gamma_{ij}^R \rangle_P + \eta_{ij}^I \langle \Gamma_{ij}^I \rangle_P \right) \\
    &- \sum_i \mu_i \left(  \langle \Gamma_{ii} \rangle_P - \frac{e^{-\beta E_i}}{\sum_{i=1}^p e^{-\beta E_i}}\right)
    - \lambda_0 \left( \langle1\rangle_P -1 \right)
    \Biggr).
\end{aligned}
\end{equation}
Solving this convex optimization problem, we find for each $\Gamma \in \mathcal{G}$, 
\begin{align}
    P[\Gamma] = \frac{
    \exp \left(-\left( \sum_{i < j}  \left( \eta_{ij}^R \langle \Gamma_{ij}^R \rangle_P + \eta_{ij}^I \langle \Gamma_{ij}^I \rangle_P \right) + \sum_{i}\mu_i \Gamma_{ii}\right) \right)
    }
    {
    \sum_{\Gamma \in \mathcal{G}}\exp \left(-\left( \sum_{i < j}  \left( \eta_{ij}^R \langle \Gamma_{ij}^R \rangle_P + \eta_{ij}^I \langle \Gamma_{ij}^I \rangle_P \right) + \sum_{i}\mu_i \Gamma_{ii}\right) \right)
    }
    \label{eq:entropyMaximizationGeneralResult}
\end{align}
We now need to determine the Lagrange multipliers by using the constraint equations from Eq.~\ref{eq:matrixConstraints}. 
Finding explicit solutions to Eq.~\ref{eq:coherenceConstraints} and \ref{eq:populationConstraints}, however, is an extremely difficult task for the complicated phase space geometry on $\Omega$.
Instead, we use the above stated closure properties for the infinitely dense grid $\mathcal{G}$ on $\Omega$.

\subsection{Determining the Lagrange Multipliers}
\paragraph{Coherence Constraints.} 
We first want to determine the $\eta_{ij}^R$ and $\eta_{ij}^I$ for the coherence constraints using the closure under real and imaginary reflection property on $\mathcal{G}$. 
We here focus on the $\eta_{ij}^R$; the analysis for the $\eta_{ij}^I$ follows in straight analogy.

The relative probability of $\Gamma$ and it's real reflected state $\Gamma^{(ij,R)}$ can be obtained from direct substitution into Eq.~\ref{eq:entropyMaximizationGeneralResult}, yielding
\begin{align}
    \frac{P[\Gamma]}{P[\Gamma^{(ij, R)}]} = \exp (-2\eta_{ij}^R \Gamma_{ij}^R)
\end{align}
We can now substitute this result into Eq.~\ref{eq:coherenceConstraints} to reduce the summation to the positive real half grid, that is $\mathcal{G}_{ij, R}^+ \equiv \{ \Gamma \in \mathcal{G}: \Gamma_{ij}^R >0\}$, which yields
\begin{equation}
\begin{aligned}
    \langle \Gamma_{ij}^R \rangle &= \sum_{\Gamma \in \mathcal{G}}P[\Gamma] \Gamma_{ij}^R = \sum_{\Gamma \in \mathcal{G}_{ij, R}^+} \Gamma_{ij}^R\left(P[\Gamma]-P[\Gamma^{(ij, R)}]\right) \\
    &= \sum_{\Gamma \in \mathcal{G}_{ij, R}^+} \Gamma_{ij}^R P[\Gamma] \left(1-\exp (-2\eta_{ij}^R \Gamma_{ij}^R)\right)
\end{aligned}
\end{equation}
for the mean real coherences. 
By requiring the mean to vanish in accordance with Eq.~\ref{eq:coherenceConstraints} and by using the same argument for the imaginary coherences, we can solve the coherence Lagrange multipliers as
\begin{align}
    \eta_{ij}^R = \eta_{ij}^I = 0.
\end{align}
This result simplifies the general result in Eq.~\ref{eq:entropyMaximizationGeneralResult} to yield
\begin{align}
    P[\Gamma] = \frac{
    \exp \left(-\sum_{i}\mu_i \Gamma_{ii}\right)
    }
    {
    \sum_{\Gamma \in \mathcal{G}}\exp \left(-\sum_{i}\mu_i \Gamma_{ii} \right)
    }
    \label{eq:entropyMaximizationSimplified}
\end{align}

\paragraph{Population Constraints.}
In order to relate the population Lagrange multipliers to the state energies and, thus, transform Eq.~\ref{eq:entropyMaximizationSimplified} into the desired form for $P[\Gamma]$ from the main text, we rewrite the $\mu_i$'s as
\begin{align}
    \mu_i = \frac{E_i}{k_\mathrm{B}\Theta_i}
\end{align}
where $\Theta_i \neq 0$ is a state-specific statistical temperature. 
We emphasize that this re-definition is simply a mathematical manipulation that shifts the Lagrange multiplier degree of freedom, $\mu_i$, into a reciprocal degree of freedom, namely $\Theta_i$.

We note that, similarly to the previous section, useful cancellation in relative probabilities can be achieved by considering two population-transferred states, \textit{i.e.}, we obtain
\begin{align}
    \frac{P[\Gamma]}{P[\Gamma^{(ij)}]} = \exp \left( -\mu_{ij} z_{ij}\right)
    \label{eq:probabilityRatioPopulationTransposition}
\end{align}
where we have defined the constants $\mu_{ij} \equiv \mu_i -\mu_j$ and the inversion coordinate $z_{ij} \equiv \Gamma_{ii}-\Gamma_{jj}$. 
Since the probabilities $P[\Gamma]$ need to be insensitive to a uniform energy shift, that is $E_i \rightarrow E_i + E_0$, the probability ratio in Eq.~\ref{eq:probabilityRatioPopulationTransposition} must also be unchanged by such a shift.
This is, however, only possible if $\mu_{ij}$ is invariant under that shift, and it, as such, can only depend on the energy difference between two states and not their absolute energies.
Using the above definition for $\mu_{ij}$, we find
\begin{align}
    \mu_{ij} = \frac{1}{k_\mathrm{B}\Theta_i} \left( E_i - \frac{\Theta_i}{\Theta_j}E_j\right)
\end{align}
and we can therefore conclude that $\Theta_i = \Theta_j\,\forall\,i,j$, \textit{i.e.}, all statistical temperatures are equal, giving the following distribution:
\begin{align}
    P[\Gamma] = \frac{e^{-\beta' \mathrm{Tr}\{\hat{H}\Gamma\}}}{\sum_{\Gamma \in \mathcal{G}}e^{-\beta' \mathrm{Tr}\{\hat{H}\Gamma\}}} = \frac{e^{-\beta' \mathrm{Tr}\{\hat{H}\Gamma\}}}{\int [d\Gamma] e^{-\beta' \mathrm{Tr}\{\hat{H}\Gamma\}}}
\end{align}
We have introduced here with $\beta' = (k_\mathrm{B}\Theta)^{-1}$ the inverse statistical temperature, and have substituted back the sum over the infinitely dense grid $\mathcal{G}$ by an integral over the uniform measure over $\Omega$, $[d\Gamma]$.

\newpage
\section{Connection to Thermodynamics} \label{sec:TempCorrespondence}
We want to establish a relationship between the temperature $\Theta$ in the $P$-ensemble, and the thermodynamic temperature, $T$, in the $\rho$-ensemble, for which we know that its equilibrium state (the Gibbs state $\rho_\mathrm{G}$) captures correctly the macroscopic thermodynamics of the system. Based on the $P$-ensemble partition function, $Z=\int [d\Gamma] e^{-\beta' \bar{H}[\Gamma]}$, the mean energy of the ensemble can be computed as
\begin{align}
\langle \bar{H}\rangle_P \equiv \int [d\Gamma] P[\Gamma]\bar{H}[\Gamma]=-\frac{\partial \ln Z}{\partial \beta'}
\label{eq:PEnsembleMeanEnergy}
\end{align}
We claim, for consistency of the $P$-ensemble with thermodynamics, that this energy has to be equal to the average energy of the Gibbs state, which in turn corresponds to the internal (thermodynamic) energy $U$ of the system, \textit{i.e.}, $U=\bar{H}[\rho_\mathrm{G}]=\langle \bar{H}\rangle_P$, and that $\beta' \equiv (k_\mathrm{B}\Theta)^{-1}$ has to be selected accordingly. Using this postulated equivalence to the internal energy, $U$, we find from Eq.~\ref{eq:PEnsembleMeanEnergy} that 
\begin{align}
    d \ln Z=-U d \beta'
    \label{eq:PEnsembleEq1}
\end{align}
Let us now introduce an auxiliary quantity $\mathcal{S} \equiv \ln Z + U \beta'$. We can clearly see, by plugging in the derived equilibrium distribution for $P[\Gamma]$, that this auxiliary quantity is exactly equal to the proposed (quantum) Shannon entropy on the density-matrix state space from the main text, $\mathcal{S}\equiv -\int [d\Gamma] P[\Gamma]\ln P[\Gamma]$, in equilibrium. The physical meaning of $\mathcal{S}$, which we will establish at the end of this section, is, however, of no relevance for the following considerations. The total differential of $\mathcal{S}$, based on the above definition, is given by 
\begin{align}
    d \mathcal{S}= d \ln Z + U d \beta' + \beta' dU,
    \label{eq:PEnsembleEq2}
\end{align}
and, by plugging in Eq.~\ref{eq:PEnsembleEq2}, we can infer that the change in internal energy can be expressed solely in terms of changes in the quantity $\mathcal{S}$ as
\begin{align}
    dU=\frac{d \mathcal{S}}{\beta'}=k_\mathrm{B}\Theta d \mathcal{S}
    \label{eq:dU_PEnsemble}
\end{align}
This result is notably independent of any thermodynamic assumptions about the $P$-ensemble and it has only taken into account the postulate that we select $\beta'$ such that $\langle \bar{H}\rangle_P =U$ in Eq.~\ref{eq:PEnsembleMeanEnergy}. \par
The First Law of thermodynamics tells us that the total change in internal energy of a (closed) system is given by $dU=TdS - p dV$, where $p$ is the pressure, $dV$ an associated volume change, and $S$ the thermodynamic entropy. We can interpret the volume as an extensive quantity that scales with the system size and with the available number of degrees of freedom. That is, upon keeping the dimensions of the system's Hilbert space fixed, we can set $dV=0$, and the First Law simplifies to
\begin{align}
    dU=TdS
    \label{eq:FistLaw}
\end{align}
Due to the equivalence of $dU$ in Eq.~\ref{eq:dU_PEnsemble} and ~\ref{eq:FistLaw} and the fact that $\Theta$ and $T$ are constants by definition, we can establish the following crucial scaling relationship between the (statistical) temperature, $\Theta$, and the (thermodynamic) temperature, $T$.  
\begin{align}
\frac{d S}{d \mathcal{S}}=\frac{k_\mathrm{B}\Theta}{T} 
\label{eq:LinearScalingTemps}
\end{align}
Not only does Eq.~\ref{eq:LinearScalingTemps} relate the temperatures, it also allows for the interpretation of the above introduced quantity $\mathcal{S}$, which we have considered initially as a measure of uncertainty over the distribution $P[\Gamma]$ as postulated by Shannon, in terms of the (thermodynamic) entropy of the system, as we find
\begin{align}
TS=k_\mathrm{B}\Theta (\mathcal{S} - \mathcal{S}_0) = k_\mathrm{B} \Theta \Bigl(\ln Z + U \beta' - \mathcal{S}_0 \Bigr)
\label{eq:TDentropy}
\end{align}
where $\mathcal{S}_0$ is just a constant of integration that can be dropped when considering entropic differences. This, indeed, shows that the quantum Shannon entropy, $\mathcal{S}$, can be interpreted in terms of the thermodynamic entropy $S$ upon scaling with $k_\mathrm{B}\Theta/T$. Furthermore, we can employ the thermodynamic definition of free energy, $F=U-TS$, to relate the (thermodynamic) free energy to the partition function $Z$. That is, we can write
\begin{align}
    F=-k_\mathrm{B}\Theta \ln Z
    \label{eq:TDfreeEnergy_1}
\end{align}
The relations in Eq.~\ref{eq:TDentropy} and ~\ref{eq:TDfreeEnergy_1} clearly prove that the $P$-ensemble partition function indeed allows us to obtain thermodynamic quantities in straight analogy to classical statistical mechanics.  

\newpage
\section{Solvation Entropy in Quantum System-Bath Composites}

We want to show here in more detail how $F[\sigma]$ can be decomposed into a enthalpic, $E[\sigma]$, and an entropic, $S[\sigma]$, contribution. The free energy for the circuit state $\sigma$ is given by the equation
\begin{equation}
\begin{aligned}
F[\sigma]&=-k_\mathrm{B}\Theta \ln \left(\int_{C_\sigma} [d\Gamma] e^{-\beta' (\bar{H}_\mathrm{S}[\Gamma]\bar{H}_\mathrm{B}[\Gamma]+\bar{V}[\Gamma])} \right) \\
&= \bar{H}_\mathrm{S}[\sigma] -k_\mathrm{B}\Theta \ln \left(\int_{C_\sigma} [d\Gamma] e^{-\beta' (\bar{H}_\mathrm{B}[\Gamma]+\bar{V}[\Gamma])} \right).
    \label{eq:solvationFE}
\end{aligned}
\end{equation}
According to Zwanzig's projection operator formalism, any generic function $\bar{O}[\Gamma]$ on phase space can be decomposed into a component projected onto a specific collective variable (here we use our quantum circuit coordinate $f[\Gamma] \equiv \mathrm{Tr_B}\{\Gamma\} = \sigma$), $\langle O \rangle_{f[\Gamma]}$, and a statistically orthogonal remainder term, $(\Delta \bar{O})_\sigma$, as
\begin{align}
\bar{O}[\Gamma]= \langle O \rangle_{f[\Gamma]}+ (\Delta \bar{O})_{f[\Gamma]}.
\label{eq:orthogonal_decomposition}
\end{align}
We wish to obtain an expression for $\langle O \rangle_{f[\Gamma]}$ as a probabilistic average over the ensemble. In order to achieve that we note that
\begin{equation}
    \begin{aligned}
        \int[d\Gamma] P[\Gamma]\bar{O}[\Gamma] \delta(f[\Gamma]-\sigma)
        &=\langle O \rangle_{\sigma} \int[d\Gamma] P[\Gamma]\delta(f[\Gamma]-\sigma)=\langle O \rangle_{\sigma} \int_{C_\sigma}[d\Gamma] P[\Gamma]
    \end{aligned}
\end{equation}
where we have expressed the $\delta$-function as a constrained integration over the subset $C_\sigma$ of the integration domain. That is, we can express $\langle O \rangle_{\sigma}$ as follows:
\begin{equation}
\begin{aligned}
    \langle O \rangle_{\sigma}&=\frac{\int_{C_{\sigma}}[d\Gamma] P[\Gamma]\bar{O}[\Gamma]}{\int_{C_{\sigma}}[d\Gamma] P[\Gamma]} \\
    &=\int_{C_{\sigma}}[d\Gamma]P[\Gamma \vert \sigma]\bar{O}[\Gamma]
\end{aligned}
\label{eq:projectionO}
\end{equation}
where $P[\Gamma \vert \sigma]$ is the conditional probability of $\Gamma$ corresponding to circuit state $\sigma$. 
We can thus use Eq.~\ref{eq:projectionO} to define the action of the projection operator onto the collective coordinate $f[\Gamma]$ as $\mathcal{P}_{\sigma}\cdot\bar{O}[\Gamma] \equiv \int_{C_{\sigma}}[d\Gamma]P[\Gamma \vert \sigma]\bar{O}[\Gamma] = \langle \bar{O}\rangle_\sigma$. 
Based on that definition, we can can easily conclude that $(1-\mathcal{P}_\sigma) \cdot \bar{O}[\Gamma] = (\Delta \bar{O})_{\sigma}$ is the corresponding (statistically) orthogonal component that captures the remainder of $\bar{O}[\Gamma]$ that is not included in $\langle O \rangle_{\sigma}$. 
Having that way defined the decomposition of $\bar{O}[\Gamma]$ in Eq.~\ref{eq:orthogonal_decomposition} from the perspective of ensemble statistics, we now apply this formalism to $\bar{V}[\Gamma]$ and $\bar{H}_\mathrm{B}[\Gamma]$ in Eq.~\ref{eq:solvationFE} to find
\begin{equation}
\begin{aligned}
    \mathcal{F}_\mathrm{solv}[\sigma]&=-k_\mathrm{B}\Theta \ln \left(\int[d\Gamma]
    e^{-\beta'
    \left(\langle \bar{H}_\mathrm{B}\rangle_{f[\Gamma]}+(\Delta \bar{H}_\mathrm{B})_{f[\Gamma]}
    +\langle \bar{V}\rangle_{f[\Gamma]}+(\Delta \bar{V})_{f[\Gamma]}
    \right)} \delta (f[\Gamma]-\sigma)
    \right)\\
    &= \langle \bar{H}_\mathrm{B}\rangle_\sigma+\langle \bar{V}\rangle_\sigma -k_\mathrm{B}\Theta \ln \left( 
    \int_{C_\sigma} [d\Gamma] e^{-\beta' 
    \left( (\Delta \bar{H}_\mathrm{B})_{\sigma}+(\Delta \bar{V})_{\sigma}\right)
    }
    \right)
\end{aligned}
\end{equation}
The enthalpic term for circuit state $\sigma$ is thus given by
\begin{equation}
    \begin{aligned}
        E[\sigma]&=\bar{H}_\mathrm{S}[\sigma]+\langle \bar{H}_\mathrm{B}\rangle_\sigma+\langle \bar{V}\rangle_\sigma \\
        &=\int_{C_\sigma}[d\Gamma] \bar{H}[\Gamma] P[\Gamma \vert \sigma]
    \end{aligned}
    \label{eq:HamMeanForce}
\end{equation}
where we have used Eq.~\ref{eq:projectionO} and the fact that $\bar{H}_\mathrm{S}[\sigma]=\int_{C_\sigma}[d\Gamma] \bar{H}_\mathrm{S}[\Gamma] P[\Gamma \vert \sigma]$. The constrained integral over the conditional probability distribution, $P[\Gamma \vert \sigma]$, reveals that $E[\sigma]$, indeed, can be interpreted as the Hamiltonian of mean force. The entropic contribution is thus given by
\begin{equation}
\begin{aligned}
    TS[\sigma]&\equiv -k_\mathrm{B}\Theta \int_{C_\sigma} [d\Gamma] P[\Gamma \vert \sigma] \ln P[\Gamma \vert \sigma]\\
    &=k_\mathrm{B}\Theta \ln \left( 
    \int_{C_\sigma} [d\Gamma] e^{-\beta' 
    \left( (\Delta \bar{H}_\mathrm{B})_{\sigma}+(\Delta \bar{V})_{\sigma}\right)
    }\right)
    .
    \label{eq:entropyQuantum}
\end{aligned}
\end{equation}
As we will show and discuss in the following subsection, this expression for the entropy behaves fundamentally different from a classical circuit-bath composite in the limit of infinitely weak coupling, \textit{i.e.}\ $\bar{V}[\Gamma] \rightarrow 0$, where we find 
\begin{align}
    TS[\sigma]=k_\mathrm{B}\Theta \ln \left( 
    \int_{C_\sigma} [d\Gamma] e^{-\beta' 
    (\Delta \bar{H}_\mathrm{B})_{\sigma}
    }
    \right).
    \label{eq:entropyQuantumInfinite}
\end{align}

\newpage
\section{Analytical Computation of the solvation free energy}
In order to analytically compute the (thermodynamic) free energy associated with a specific single-system density matrix, $F[\sigma]=-k_\mathrm{B}\Theta \ln Z[\sigma]$, we need to compute the constrained partition function along a fixed $\sigma$, given by
\begin{equation}
\begin{aligned}
    Z[\sigma]&=\int_{C_\sigma} [d\Gamma] \exp(-\beta' \bar{H}[\Gamma]) \\
    & =\int [d\Gamma] \delta(\sigma-\mathrm{Tr_B}\{\Gamma\})\exp(-\beta' \bar{H}[\Gamma]),
    \label{eq:constrainedPartitionFunction}
\end{aligned}
\end{equation}
where we have used that the constrained integration over the set of microstates $\Gamma$ that corresponds to a particular macrostate $\sigma$, $C_\mathrm{\sigma} \equiv \{\Gamma \in \mathcal{H}: \mathrm{Tr_B}\{\Gamma\}=\sigma\}$, can be written as a $\delta$-function. In noting that, for the system-bath composite, $\bar{H}[\Gamma]=\bar{H}_\mathrm{S}[\Gamma]+\bar{H}_\mathrm{B}[\Gamma]+\bar{V}[\Gamma]$ and $\bar{H}_\mathrm{S}[\Gamma]=\bar{H}_\mathrm{S}[\sigma]$ because $\hat{H}_\mathrm{S} \in \mathcal{H}_\mathrm{S}$, we can separate $Z[\sigma]$ into a (trivial) system contribution and a non-trivial component, here denoted as $Z'[\sigma]$, which encapsulated the effect of the bath onto the ensemble properties of the system, with
\begin{equation}
\begin{gathered}
    Z[\sigma]=\exp (-\beta' \bar{H}_\mathrm{S}[\sigma]) Q'[\sigma] \\
    \text{and} \quad Z'[\sigma] =\int [d\Gamma] \delta(\sigma-\mathrm{Tr_B}\{\Gamma\})\exp(-\beta' (\bar{H}_\mathrm{B}[\Gamma]+\bar{V}[\Gamma])).
\label{eq:Qprime}
\end{gathered}
\end{equation}
Finding analytical solutions to Eq.~\ref{eq:Qprime}, is equivalent to computing the solvation free energy, $\mathcal{F}_\mathrm{solv}[\sigma] \equiv -k_\mathrm{B}\Theta \ln Z'[\sigma]$, which in turn gives us access to the free energy by $F[\sigma]= \bar{H}_\mathrm{S}[\sigma]+\mathcal{F}_\mathrm{solv}[\sigma]$ and therefore to the thermodynamic driving forces associated with state $\sigma$ that we are interested in eventually. \par
The integral in Eq.~\ref{eq:Qprime}, however, is highly non-trivial to solve for the complex phase space geometry in the quantum domain:
\begin{enumerate}[(i)]

\item The integration measure $[d\Gamma]$ denotes the uniform measure over all Hermitian operators on $\mathcal{H} = \mathcal{H}_\mathrm{S} \otimes \mathcal{H}_\mathrm{B}$ of the set $S_\Gamma \equiv \{\Gamma \in \mathbb{C}^{NM} \times \mathbb{C}^{NM}: \Gamma = \Gamma^\dagger, \Gamma >0, \mathrm{Tr}\{\Gamma\}=\mathrm{Tr}\{\Gamma^2\}=1\}$, where we have defined $\mathrm{dim}(\mathcal{H}_\mathrm{S})\equiv N$ and $\mathrm{dim}(\mathcal{H}_\mathrm{B}) \equiv M$. Since states $\Gamma$ of the system-bath composite are pure (\textit{i.e.} we can write $\Gamma=\vert \Psi \rangle \langle \Psi \vert$ with $\vert \Psi \rangle \in \mathbb{C}^{NM}$) the number of degrees of freedom in the measure $[d\Gamma]$, typically a product over all matrix elements subject to the constraints specified by $S_\Gamma$, can be dramatically reduced. If we represent the system-bath wavefunction in the product basis, $\{\vert i \rangle \otimes \vert j \rangle\}$, with $\{\vert i \rangle \in \mathcal{H}_\mathrm{S}\}$ and $\{\vert j \rangle \in \mathcal{H}_\mathrm{B}\}$ being complete orthonormal basis sets on the system and bath Hilbert spaces, respectively, we can write $\vert \Psi \rangle$ as
\begin{align}
\vert \Psi \rangle =\sum_{i=1}^N \sum_{j=1}^M c_{ij}\vert i \rangle \otimes \vert j \rangle.
\label{eq:waveFunctionSysBath}
\end{align}
This representation allows us to cast all information contained within $\vert \Psi \rangle$ into a complex $N \times M$ coefficient matrix, $X$, with $X_{ij}\equiv c_{ij}$. It can easily be verified that the normalization constraint on the wavefunction, $\langle \Psi \vert \Psi \rangle=1$, is given by the (squared) Frobenius norm of matrix $X$, \textit{i.e.}\ 
\begin{align}
\langle \Psi \vert \Psi \rangle=\Vert X\Vert_\mathrm{F}^2=\mathrm{Tr}\{X X^\dagger\}=\mathrm{Tr}\{X^\dagger X\}=1.
\label{eq:normConstraint}
\end{align}
The uniform measure over $S_\Gamma$, $[d\Gamma]$, can therefore be transformed into the uniform measure over $S_X :=\{X \in \mathbb{C}^{N \times M}\}$, $[dX]$, weighted by the $\delta$-function constraint $\delta(\mathrm{X X^\dagger} - 1)$.

\item The change in the integration variable from $\Gamma$ to $X$ also necessitates a transformation of subset $C_\sigma$ in Eq.~\ref{eq:Qprime}. It can be confirmed easily that the operation of tracing over the bath degrees of freedom, $\mathrm{Tr_B}\{\Gamma\}$, can be written as the matrix product $XX^\dagger$. We can thus write the $\delta$-function  constraint in Eq.~\ref{eq:Qprime} as 
\begin{align}
\delta(\mathrm{Tr_B}\{\Gamma\}-\sigma)=\delta(XX^\dagger -\sigma)
\label{eq:tracingConstraintX}
\end{align}

\item We note that $\bar{H}_\mathrm{B}[\Gamma] \equiv \mathrm{Tr}\{\hat{H}_\mathrm{B}\Gamma\}=\mathrm{Tr}\{\hat{H}_\mathrm{B} \gamma\}$ where $\gamma = \mathrm{Tr_S}\{\Gamma\}$ is the reduced bath density matrix since $\hat{H}_\mathrm{B} \in \mathcal{H}_\mathrm{B}$ is acting on the bath Hilbert space only. The trace over the system Hilbert space can, analogously to the bath trace, conveniently be obtained from $X$ by computing $\gamma = X^\dagger X$. This allows us to rewrite the bath energy, $\bar{H}_\mathrm{B}[\Gamma]$, as
\begin{align}
    \bar{H}_\mathrm{B}[\Gamma]=\mathrm{Tr}\{H_\mathrm{B} X^\dagger X\},
    \label{eq:bathEnergyX}
\end{align}
where $H_\mathrm{B}$ is the $M \times M$-dimensional matrix representation of $\hat{H}_\mathrm{B}$ in the basis $\{\vert j \rangle\}$. \par   
Such a change in notation is not possible for $\hat{V}$ as $\hat{V}\in \mathcal{H}$ is not exclusively confined to either of the two subspaces $\mathcal{H}_\mathrm{S}$ or $\mathcal{H}_\mathrm{B}$. However, we assume that system and bath are only weakly coupled to each other, \textit{i.e.} 
\begin{align}
    \bar{V}[\Gamma] \ll \bar{H}_\mathrm{B}[\Gamma], \bar{H}_\mathrm{S}[\Gamma].
    \label{eq:weakCoupling}
\end{align}
In neglecting $\bar{V}[\Gamma]$ we assume that the equilibrium distribution is essentially unaffected by the coupling, which appears to be a sensible approximation as the Lamb-shift term in conventional approaches with perturbative master equations is vanishingly small and the Gibbs state remains unchanged upon marginal changes to the coupling. While acknowledging that, strictly speaking, this approximation only allows for a treatment of system-bath coupling up to zeroth order, we emphasize that the analytical solutions we are outlining in the following sections, except for this weak-coupling assumption, are exact. 

\end{enumerate}

Equipped with these preliminaries, we now want to solve Eq.~\ref{eq:Qprime} for different bath conditions.

\subsection{Bath with Degenerate Energy Levels}
We start our considerations by taking the bath to be a collection of $M$ degenerate states with the same energy $E_0$. In this limit, each bath state occurs with equal a priori probabilities and the partition function $Z'[\sigma]$ can be rationalized in the Boltzmann-sense as a number of microstates corresponding to a particular macrostate $\sigma$, \textit{i.e.} 
\begin{equation}
\begin{aligned}
    &Z'[\sigma]=\int [d\Gamma] \delta(\sigma-\mathrm{Tr_B}\{\Gamma\}) \\
    \rightarrow \quad &Z'[\sigma]=\int_{\mathbb{C}^{N\times M}} [dX] \delta(1-\mathrm{Tr}\{XX^\dagger\})\delta(\sigma-XX^\dagger),
\label{eq:degenerateQprime}
\end{aligned}
\end{equation}
where we have introduced the change of notation as discussed above. We have also dropped the constant energy factor $\exp(-\beta'E_0)$ as it does not affect the ensemble statistics. In order to solve Eq.~\ref{eq:degenerateQprime}, we first perform the substitution $X \rightarrow \tilde{X}$ with 
\begin{align}
    X\equiv \sqrt{\sigma} \tilde{X}
    \label{eq:substitution}
\end{align}
and $\tilde{X} \in \mathbb{C}^{N \times M}$. Note that $\sqrt{\sigma}$ is well-defined as the density matrix is positive semidefinite and, as such, has non-negative eigenvalues. The Jacobian of this transformation is given by 
\begin{align}
[dX]=\det[\sigma]^M [d\tilde{X}]
\label{eq:Jacobian}
\end{align}
as following from Proposition 3.4 in Ref.\ \cite{Zhang2015}. Upon this change of variable, we also introduce a scaling factor into $\delta(\sigma-XX^\dagger)$, determined by $\sigma$, which needs to be accounted for. We can write 
\begin{equation}
\begin{aligned}
    \delta(\sigma-XX^\dagger)&= \delta(\sqrt{\sigma}(\mathbbm{1}_{N}-\tilde{X}\tilde{X}^\dagger)\sqrt{\sigma}) \\
    &=\det[\sigma]^{-N}\delta(\mathbbm{1}_{N}-\tilde{X}\tilde{X}^\dagger),
    \label{eq:DeltaSubstitution}
\end{aligned}
\end{equation}
using Proposition 3.11 in Ref.\ \cite{Zhang2021}. If we now perform the substitution $X \rightarrow \tilde{X}$ and use Eqs.~\ref{eq:substitution}, ~\ref{eq:Jacobian}, and ~\ref{eq:DeltaSubstitution} in the integral for $Q'[\sigma]$ as well as the equality $\sigma=XX^\dagger$, we obtain 
\begin{equation}
\begin{aligned}
    Z'[\sigma]&=\det[\sigma]^{M-N}\int_{\mathbb{C}^{N\times M}}[d\tilde{X}]\delta(1-\mathrm{Tr}\{\sigma\})\delta(\mathbbm{1}_N-\tilde{X}\tilde{X}^\dagger)\\
    &=\det[\sigma]^{M-N}\delta(1-\mathrm{Tr}\{\sigma\})\int_{\mathbb{C}^{N\times M}}[d\tilde{X}]\delta(\mathbbm{1}_N-\tilde{X}\tilde{X}^\dagger)\\
    &=\det[\sigma]^{M-N}\int_{\mathbb{C}^{N\times M}}[d\tilde{X}]\delta(\mathbbm{1}_N-\tilde{X}\tilde{X}^\dagger).
    \label{eq:degenerateQprime1}
\end{aligned}
\end{equation}
We have used, in the second line, the fact that $\delta(1-\mathrm{Tr}\{\sigma\})$ is independent of the integration variable $\tilde{X}$. Evidently, this constraint, originating from the normalization of the system-bath wave function $\delta(1-\mathrm{Tr}\{XX^\dagger\})$, mathematically reduces to the postulate that the system density matrix needs to have unit trace. This postulate can naturally be understood as the normalization condition on $\sigma$ since the probabilities of being in any eigenstate, upon decomposition of $\sigma$ into those eigenstates, have to add up to one. Albeit not having been clearly specified before in our attempt to solve Eq.~\ref{eq:degenerateQprime}, we drop this constraint in the third line by redundancy, acknowledging that a physically reasonable density matrix $\sigma$ will always have the property $\mathrm{Tr}\{\sigma\}=1$ by construction. \par

In Eq.~\ref{eq:degenerateQprime1}, we have shown how $Z'[\sigma]$ can be decomposed into a $\sigma$-dependent term and a term that, in a non-trivial way, reflects the geometry of the underlying phase space upon the operation of computing the partial trace over the bath. While $\tilde{X}$ cannot be rationalized intuitively anymore in terms of the system-bath state, $X$, we want to mathematically describe the topological space specified by $\delta(\mathbbm{1}_N-\tilde{X}\tilde{X}^\dagger)$ and find a closed form solution to the remainder integral. \\
In order to do that, we note that the integration measure is invariant under the dagger operation $\tilde{X} \in \mathbb{C}^{N \times M} \rightarrow \tilde{X}^\dagger \in \mathbb{C}^{M \times N}$, which allows us to rewrite Eq.~\ref{eq:degenerateQprime1} as
\begin{align}
    Z'[\sigma]=\det[\sigma]^{M-N}\int_{\mathbb{C}^{M \times N}} [d\tilde{X}]\delta(\mathbbm{1}_N-\tilde{X}^\dagger \tilde{X}).
    \label{eq:degenerateQprime2}
\end{align}
The topological manifold spanned by $\delta(\mathbbm{1}_N-\tilde{X}^\dagger \tilde{X})$ in Eq.~\ref{eq:degenerateQprime2} is referred to as the (complex) Stiefel manifold, defined as the set $\mathcal{V}_N(\mathbb{C}^M) \equiv \{A \in \mathbb{C}^{M \times N}: A^\dagger A=\mathbbm{1}_N\}$, and matrices $\tilde{X} \in \mathcal{V}_N(\mathbb{C}^M)$ are called semi-unitary. Integrals over the Stiefel manifold have been studied in literature, but only with respect to the Haar measure on that manifold. For the (complex) Stiefel manifold, this measure is given by $[\tilde{X}^\dagger d\tilde{X}]$ \cite{Chikuse2003, Zhang2015, Zhang2021} and, according to Remark 4.3 in Ref.\ \cite{Zhang2021}, is related to the (unnormalized) Haar measure on $\mathbb{C}^{M\times N}$, $[d\tilde{X}]$, by
\begin{align}
    [\tilde{X}^\dagger d\tilde{X}]=2^N [d\tilde{X}]
    \label{eq:haarMeasureStiefel}
\end{align}
That is, we rewrite Eq.~\ref{eq:degenerateQprime2} as
\begin{equation}
\begin{aligned}
    Z'[\sigma]&=2^{-N}\det[\sigma]^{M-N}\int_{\mathbb{C}^{M \times N}} [\tilde{X}^\dagger d\tilde{X}]\delta(\mathbbm{1}_N-\tilde{X}^\dagger \tilde{X})\\
    &=2^{-N}\det[\sigma]^{M-N}\int_{V_N(\mathbb{C}^M)} [\tilde{X}^\dagger d\tilde{X}]\\
    &=2^{-N}\det[\sigma]^{M-N}\mathrm{Vol(V_N(\mathbb{C}^M))},
    \label{eq:degenerateQprime3}
\end{aligned}
\end{equation}
where we have defined $\mathrm{Vol(\mathcal{V}_N(\mathbb{C}^M))} \equiv \int_{\mathcal{V}_N(\mathbb{C}^M)} [\tilde{X}^\dagger d\tilde{X}]$ as the volume over the Stiefel manifold $V_N(\mathbb{C}^M)$. This volume is given by
\begin{align}
    \mathrm{Vol(\mathcal{V}_N(\mathbb{C}^M))}=\frac{2^N \pi^{\frac{1}{2}N(2M-N+1)}}{\prod_{k=1}^N(M-k)!},
    \label{eq:StiefelVolume}
\end{align}
which concludes the computation of the partition function $Q'[\sigma]$ in Eq.~\ref{eq:degenerateQprime}. Using the relation $\mathcal{F}_\mathrm{solv}[\sigma]=-k_\mathrm{B}\Theta \ln Z'[\sigma]$ for the solvation free energy, we obtain
\begin{align}
\mathcal{F}_\mathrm{solv}[\sigma]=-k_\mathrm{B}\Theta \ln \det[\sigma]^{M-N}+k_\mathrm{B}\Theta\ln \Bigl( 2^{-N} \mathrm{Vol}(\mathcal{V}_N(\mathbb{C}^M)) \Bigr).
\label{eq:solvationFEdegenerate}
\end{align}

\subsection{Thermal Bath with Distinct Energy Levels}
We now consider the case for which the $M$ bath states have distinct, \textit{i.e.} non-degenerate energies. The partition function $Z'[\sigma]$ that we now have to solve is given by 
\begin{equation}
\begin{aligned}
    &Z'[\sigma]=\int [d\Gamma] \delta(\sigma-\mathrm{Tr_B}\{\Gamma\})\exp(-\beta'\bar{H}_\mathrm{B}[\Gamma]) \\
    \rightarrow \quad &Z'[\sigma]=\int_{\mathbb{C}^{N\times M}} [dX] \delta(1-\mathrm{Tr}\{XX^\dagger\})\delta(\sigma-XX^\dagger) \exp(-\beta'\mathrm{Tr}\{H_\mathrm{B}X^\dagger X\}).
\label{eq:distinctQprime}
\end{aligned}
\end{equation}
As seen for the degenerate bath, we perform the substitution $X \rightarrow \tilde{X}$ with $X \equiv \sqrt{\sigma}\tilde{X}$ (see Eq.~\ref{eq:substitution}). By the same argument that had been used in Eq.~\ref{eq:degenerateQprime1}, this substitution leads to 
\begin{align}
   Z'[\sigma]=\det[\sigma]^{M-N}\int_{\mathbb{C}^{N\times M}}[d\tilde{X}]\delta(\mathbbm{1}_N-\tilde{X}\tilde{X}^\dagger)\exp(-\beta'\mathrm{Tr}\{\tilde{X}\sigma \tilde{X}^\dagger H_\mathrm{B}\}).
   \label{eq:distQprimeSubs}
\end{align}
In order to make progress on the remaining integral, we again want to perform a change of measure. This time, however, we want to relate $[d\tilde{X}]$ to the normalized Haar measure over $V_N(\mathbb{C}^M)$, $(d\tilde{X})$, which is defined such that
\begin{align}
    1 = \int_{\mathcal{V}_N(\mathbb{C}^M)} (d\tilde{X}),
    \label{eq:normHaarMeasure}
\end{align}
and therefore related to the unnormalized Haar measure over that manifold, $[\tilde{X}^\dagger d\tilde{X}]$, by $[\tilde{X}^\dagger d\tilde{X}]=\mathrm{Vol(\mathcal{V}_N(\mathbb{C}^M))}(d\tilde{X})$. Combining this with Eq.~\ref{eq:haarMeasureStiefel}, we find
\begin{align}
    [d\tilde{X}]=2^{-N}\mathrm{Vol(\mathcal{V}_N(\mathbb{C}^M))}(d\tilde{X}),
    \label{eq:measureChangeNorm}
\end{align}
and Eq.~\ref{eq:distQprimeSubs} can be written as an integral over the Haar probability measure on the complex Stiefel manifold:
\begin{equation}
\begin{aligned}
    Z'[\sigma]&=2^{-N}\mathrm{Vol(V_N(\mathbb{C}^M))}\det[\sigma]^{M-N}J[\sigma]\\
    \mathrm{with}\quad & J[\sigma] \equiv \int_{V_N(\mathbb{C}^M)}(d\tilde{X})\exp(-\beta'\mathrm{Tr}\{\tilde{X}\sigma \tilde{X}^\dagger H_\mathrm{B}\})
\label{eq:distQ_HZICform}
\end{aligned}
\end{equation}
Formally, the integral $J[\sigma]$ in Eq.~\ref{eq:distQ_HZICform} corresponds to the definition of the heterogeneous hypergeometric function $ \tensor[_0]{F}{_0}(A,B)$ of two Hermitian matrix arguments $A$ ($N \times N$ matrix) and $B$ ($M \times M$ matrix), defined as 
\begin{align}
    \tensor[_0]{F}{_0}(A,B) = \int_{V_N(\mathbb{C}^M)}(d\tilde{X})\exp(-\beta'\mathrm{Tr}\{\tilde{X}A \tilde{X}^\dagger B\}),
    \label{eq:hypergeomFunction}
\end{align}
where we can choose $A \equiv \sigma$ and $B \equiv -\beta'H_\mathrm{B}$. This integral over the Stiefel manifold is hard to evaluate. Analytical approximations for $ \tensor[_0]{F}{_0}(A,B)$ can be obtained by Laplace approximations as shown by Butler and Wood \cite{Butler2005}. 
In an attempt to find a closed-form analytical solution, we first transform the integration over the Stiefel manifold $V_N(\mathbb{C}^M)$ in Eq.~\ref{eq:hypergeomFunction} into an integration over the unitary group $U_M \equiv \{Y \in C^{M\times M}: Y Y^\dagger = Y^\dagger Y=\mathbbm{1}_M\}$ and into the normalized Haar measure over that group, $(dY)$, with $M > N$. First, we define the $M \times M$ Hermitian block matrix
\begin{align}
\setlength\arraycolsep{7pt}
\begingroup
    \sigma' \equiv 
    \begin{pmatrix}
   \sigma & \bigzero_{N \times (M-N)} \\
   \bigzero_{(M-N) \times N} & \bigzero_{(M-N) \times (M-N)}
 \end{pmatrix},
 \endgroup
 \label{eq:newSigma}
\end{align}
where $\sigma$ is the $N \times N$-dimensional system density matrix, and $\bigzero_{i\times j}$ are $i \times j$-dimensional null matrices. According to Lemma 1 and Theorem 1 in Ref.\ \cite{Shimizu2021}, this definition enables us to write Eq.~\ref{eq:distQ_HZICform} as
\begin{equation}
    \begin{aligned}
        J[\sigma] = \int_{V_N(\mathbb{C}^M)}(d\tilde{X})\exp(-\beta'\mathrm{Tr}\{\tilde{X}\sigma \tilde{X}^\dagger H_\mathrm{B}\}) = \int_{U_M} (dY)\exp(-\beta'\mathrm{Tr}\{Y \sigma' Y^\dagger H_\mathrm{B}\}).
            \label{eq:distQ_HZIC2}
    \end{aligned}
\end{equation}
The authors in Ref.\ \cite{Shimizu2021} only show this implication for the transformation of integrals from the real Stiefel manifold to the orthogonal group (the real analogue to the complex Stiefel manifold and the unitary group), but the Hermiticity of matrices involved (\textit{i.e.}\ the condition that thos matrices have real eigenvalues) naturally extends this proof to the transformation from the complex Stiefel manifold to the unitary group. \par
It can be noted that the integral over $U_M$ in Eq.~\ref{eq:distQ_HZIC2} is the well-studied \textit{Harish-Chandra-Itzykson-Zuber (HCIZ) integral} that has a closed-form solution \cite{McSwiggen2021}. Let $A$, $B$ be Hermitian $M \times M$ matrices with eigenvalues $\lambda_1(A) \leq \lambda_2(A) \leq ... \leq \lambda_M(A)$ and $\lambda_1(B) \leq \lambda_2(B) \leq ... \leq \lambda_M(B)$. If $\alpha$ is a non-zero parameter, then 
\begin{equation}
    \begin{aligned}
        \int_{U_M} (dY)\exp (\alpha &\mathrm{Tr}\{AYBY^\dagger\})=c_M \frac{\det \Bigl[ e^{\alpha \lambda_i(A)\lambda_j(B)}\Bigr]_{1 \leq i, j \leq M}}{\alpha^{\frac{M^2-M}{2}}\Delta(A)\Delta(B)} \\[18pt]
        \text{with} \quad  & \Delta(A) \equiv \prod_{1 \leq i < j \leq M} \left( \lambda_i(A)-\lambda_j(B)\right) \\
        &\Delta(B) \equiv \prod_{1 \leq i < j \leq M} \left( \lambda_i(B)-\lambda_j(B)\right) \\
        &c_M =\prod_{i=1}^M i!.
    \label{eq:HCIZ_general}
    \end{aligned}
\end{equation}
In order to solve Eq.~\ref{eq:distQ_HZIC2}, we can use this HCIZ integral formula and set $A=H_\mathrm{B}$, $B=\sigma'$ and $\alpha=-\beta'$. \\
We note, however, that $\lambda_1(\sigma')=\lambda_2(\sigma')=...=\lambda_{M-N}(\sigma')=0$, \textit{i.e.}\ the $(M-N)$ smallest eigenvalues of $\sigma'$ are zero. That is, the determinant of the matrix $\left[ e^{-\beta' \lambda_i(H_\mathrm{B})\lambda_j(\sigma')}\right]_{i,j}$ vanishes and so does $\Delta(\sigma')$. We resolve this dilemma by repetitive usage of L'Hôpital's rule, which yields
\begin{align}
    \frac{\det \Bigl[ e^{-\beta' \lambda_i(H_\mathrm{B})\lambda_j(\sigma')}\Bigr]_{1 \leq i, j \leq M}}{\Delta(\sigma')} \rightarrow 
    \left( \prod_{i=1}^{M-N-1} \frac{(-1)^i}{i!}\right)
    \frac{\det [I(\sigma)]}{\Delta(\sigma) \left(\prod_{i=1}^N \lambda_i(\sigma) \right)^{M-N}}
    \label{eq:HCIZ_LHopital}
\end{align}
where we have defined the real $M\times M$ matrix
\begin{align}
    I(\sigma)=\Bigl[ 1, (\beta'\lambda_i(H_\mathrm{B})), ..., (\beta'\lambda_i(H_\mathrm{B}))^{M-N-1}, e^{-\beta' \lambda_1(\sigma)\lambda_i(H_\mathrm{B})},...,e^{-\beta' \lambda_N(\sigma)\lambda_i(H_\mathrm{B})}\Bigr]_{i=1}^M
    \label{eq:interactionMatI}
\end{align}
and used the fact that $\lambda_{M-N+1}(\sigma')=\lambda_1(\sigma), ..., \lambda_M(\sigma')=\lambda_N(\sigma)$. In furthermore recognizing that $\left(\prod_{i=1}^N \lambda_i(\sigma) \right)^{M-N}=\det[\sigma]^{M-N}$ and by combining Eqs.~\ref{eq:distQ_HZIC2}, ~\ref{eq:HCIZ_general}, and ~\ref{eq:HCIZ_LHopital}, we can obtain a closed-from solution to the constrained partition function $Z'[\sigma]$ in Eq.~\ref{eq:distQ_HZICform}, and consequently also to the solvation free energy via $\mathcal{F}_\mathrm{solv}[\sigma]=-k_\mathrm{B}\Theta \ln Z'[\sigma]$ as
\begin{align}
        \mathcal{F}_\mathrm{solv}[\sigma]=-k_\mathrm{B}\Theta \Biggl[  \ln \left( \frac{\vert \det [I(\sigma)] \vert}{\Delta(\sigma)}  \right) +\ln \left( \frac{2^{-N} \mathrm{Vol} \left(\mathcal{V}_N(\mathbb{C}^M)\right)\prod_{i={M-N}}^M i!}{\beta'^{\frac{M(M-1)}{2}}\Delta(H_\mathrm{B})} \right)\Biggr].
        \label{eq:solvationFEdistinct}
\end{align}
We have used in this derivation the fact that the spectrum $\Lambda(H_\mathrm{B})$ is non-degenerate. While Eq.~\ref{eq:solvationFEdistinct} is a consequence of that assumption, we acknowledge that this condition is not strictly necessary for deriving closed-form solutions to $\mathcal{F}_\mathrm{solv}[\sigma]$ as we have shown in Eq.~\ref{eq:HCIZ_LHopital} how eigenvalue degeneracies can be dealt with by repetitive use of L'Hôpital's rule. The same applies to degeneracies in the spectrum of $\sigma$.    

\subsection{Obtaining the Entropy from the Solvation Free Energy}
In the previous two subsections, we have essentially derived analytical closed form expressions for the (thermodynamic) free energy, $F[\sigma]$, associated with a system state $\sigma$ for two different bath conditions. In fact, we have computed $\mathcal{F}_\mathrm{solv}[\sigma]$, which can easily be related to $F[\sigma]$ by
\begin{align}
    F[\sigma]=\bar{H}_\mathrm{S}[\sigma]+\mathcal{F}_\mathrm{solv}[\sigma]=-k_\mathrm{B}\Theta \ln Z[\sigma].
    \label{eq:FreeEnergyRelation}
\end{align}
It is natural to assume that the thermodynamic free energy can be separated into an entropic and enthalpic contribution according to
\begin{align}
    F[\sigma]=E[\sigma]-TS[\sigma].
    \label{eq:TDfreeEnergy}
\end{align}
The enthalpic contribution, $E[\sigma]$, can be rationalized in classical analogy as the Hamiltonian of mean force along for a fixed value of the collective variable $\sigma$. It is given by
\begin{align}
    E[\sigma]=\int_{C_\sigma} [d\Gamma] P[\Gamma \vert \sigma] \bar{H}[\Gamma]=-\left( \frac{\partial \ln Z[\sigma]}{\partial \beta'}\right),
    \label{eq:HamMeanForce1}
\end{align}
where 
\begin{align}
    P[\Gamma \vert \sigma]=\frac{e^{-\beta' \bar{H}[\Gamma]}}{Z[\sigma]}
    \label{eq:conditionalProb}
\end{align}
is the conditional probability distribution of being in a microstate $\Gamma$ given a macrostate $\sigma$. In combining Eqs.~\ref{eq:FreeEnergyRelation}, \ref{eq:TDfreeEnergy}, and \ref{eq:HamMeanForce1}, we can obtain an expression for the thermodynamic entropy, $S[\sigma]$, based on the analytical solution for $Z[\sigma]$. 

\paragraph{Weakly coupled degenerate Bath.}
For the degenerate bath, we find $E[\sigma]=\bar{H}_\mathrm{S}[\sigma]$, and therefore $S[\sigma]=-\frac{1}{T}\mathcal{F}_\mathrm{solv}[\sigma]$, with $\mathcal{F}_\mathrm{solv}[\sigma]$ being specified in Eq.~\ref{eq:solvationFEdegenerate}. We find that 
\begin{align}
    TS[\sigma]=k_\mathrm{B}\Theta \ln \det[\sigma]^{M-N} + \mathrm{const.}, 
    \label{eq:EntropyDegenerate}
\end{align}
where we have dropped terms independent of $\sigma$. We note that the latter term vanishes as soon as we consider entropy differences between different system states $\sigma$. 

\paragraph{Weakly coupled non-degenerate bath.} For the thermal bath, the Hamiltonian of mean force, however, does not simply reduce to $\bar{H}_\mathrm{S}[\sigma]$. By using $\ln Z[\sigma]=-\beta'\bar{H}_\mathrm{S}[\sigma]+\ln Q'[\sigma]$ and Eq.~\ref{eq:HamMeanForce}, we can compute 
\begin{equation}
    \begin{aligned}
        E[\sigma]&=\bar{H}_\mathrm{S}[\sigma]-\frac{\partial }{\partial \beta'} 
        \Biggl[ 
         \ln \left( \frac{\vert \det [I(\sigma)]\vert}{\Delta(\sigma)}  \right) +\ln \left( \frac{2^{-N} \mathrm{Vol} \left(\mathcal{V}_N(\mathbb{C}^M)\right)\prod_{i=M-N}^M i!}{\beta'^{\frac{M(M-1)}{2}}\Delta(H_\mathrm{B})} \right)
        \Biggr] \\
        &=\bar{H}_\mathrm{S}[\sigma] - \frac{1}{\vert \det[I(\sigma)]\vert} \cdot \frac{\partial \vert \det [I(\sigma)]\vert}{\partial \beta'} + \frac{M(M-1)}{2\beta'} \\
        & = \bar{H}_\mathrm{S}[\sigma] - \frac{\mathrm{sgn}(\det[I(\sigma)])}{\mathrm{sgn}(\det[I(\sigma)]) \det[I(\sigma)]} \cdot \frac{\partial \det [I(\sigma)]}{\partial \beta'} + \frac{M(M-1)}{2\beta'} \\
        & = \bar{H}_\mathrm{S}[\sigma] - \frac{1}{\det[I(\sigma)]} \cdot \frac{\partial \det [I(\sigma)]}{\partial \beta'} + \frac{M(M-1)}{2\beta'}, \\
    \end{aligned}
    \label{eq:energyThermalBath1}
\end{equation}
where we have used the chain rule and $f'(x)= \mathrm{sgn}(x)$ for the absolute value function $f(x)=\vert x \vert$.
In order to compute the partial derivative of the determinant in Eq.~\ref{eq:energyThermalBath1}, we apply Jacobi's formula and obtain
\begin{equation}
    \begin{gathered}
        E[\sigma]=\bar{H}_\mathrm{S}[\sigma] - \mathrm{Tr}\Bigl\{ I(\sigma)^{-1}\cdot I(\sigma)'\Bigr\} + \frac{M(M-1)}{2\beta'} \\
        \text{with} \quad   I(\sigma)' \equiv \frac{\partial I(\sigma)}{\partial \beta'}=\\
        \Bigl[0, \lambda_i, \frac{2(\beta'\lambda_i)^2}{\beta'}, ..., \frac{(M-N-1))(\beta'\lambda_i)^{M-N-1}}{\beta'},  -\lambda_i\eta_1e^{-\beta' \lambda_i\eta_1},...,-\lambda_i\eta_Ne^{-\beta' \lambda_i\eta_N}\Bigr]_{i=1}^M,
    \label{eq:energyThermalBath2}
    \end{gathered}
\end{equation}
where $I(\sigma)^{-1}$ is the (unit-less) inverse of matrix $I(\sigma)$ in Eq.~\ref{eq:interactionMatI} and $I(\sigma)'$ its scalar derivative with respect to $\beta'$ (with units energy). That is, by combining Eq.~\ref{eq:TDfreeEnergy} with Eq.~\ref{eq:energyThermalBath2} and the expression for the free energy associate with a non-degenerate bath, Eq.~\ref{eq:solvationFEdistinct}, we obtain the entropy
\begin{align}
    TS[\sigma]=k_\mathrm{B}\Theta \ln \left( \frac{\vert \det [I(\sigma)]\vert}{\Delta(\sigma)}  \right)-\mathrm{Tr}\Bigl\{ I(\sigma)^{-1}\cdot I(\sigma)'\Bigr\}+\mathrm{const.},
    \label{eq:EntropyDistinct}
\end{align}
where we have dropped all terms that are independent of $\sigma$. We can use Eq.~\ref{eq:EntropyDistinct} to compute entropy differences for arbitrary transformations $\sigma_\mathrm{i} \rightarrow \sigma_\mathrm{f}$ as done in the main text.

\newpage
\section{Markov Chain Monte Carlo Simulations}
In order to numerically generate samples from $P[\sigma]$ so as to validate our analytical results, we perform Markov Chain Monte Carlo (MCMC) simulations according to the Metropolis-Hastings algorithm. 
Rather than sampling from the state space of (potentially mixed) system density matrices $\sigma$, however, we rather use MCMC to sample from $P[\Gamma]$ on the state space of pure system-bath states $\Gamma$, and then obtain the lower-dimensional system coordinate of interest, $\sigma = \mathrm{Tr_B}\{\Gamma\}$.

Let $\hat{H}, \Gamma \in \mathcal{H}$ be the (known) system-bath Hamiltonian and the system-bath density operator, respectively, with $\mathrm{dim}(\mathcal{H})=NM$. 
Due to the purity of states $\Gamma$, we can write $\mathrm{Tr}\{\hat{H}\Gamma\} = \langle \Psi \vert \hat{H} \vert \Psi \rangle$, where $\Gamma = \vert \Psi \rangle \langle \Psi \vert$ and $\vert \Psi \rangle \in \mathbb{C}^{NM}$ being a system-bath wave function with normalization constraint $\langle \Psi \vert \Psi \rangle = 1$. 
Each $\vert \Psi \rangle = (\psi_1, \psi_2, ..., \psi_{NM})^T$ with $\psi_i = \Re[\psi_i] + i\Im[\psi_i] \in \mathbb{C}$ has $2NM -1$ real degrees of freedom. 
That is, the problem of sampling complex system-bath wave functions reduces to the problem of probing the surface of a real $(2NM-1)$-sphere $S^{2NM-1} \equiv \{ \mathbf{x} \in \mathbb{R}^{2NM} :\left\| \mathbf{x}\right\|=1 \}$ and setting $\Re[\psi_i] = x_i$, $\Im[\psi_i] = x_{NM+i}$. 
One strategy to generate ergodic Markov chains $\{\Psi_0, \Psi_1, ...\}$ on the state space of $\Gamma$ therefore is to draw states $\vert \Psi_i \rangle$ uniformly from $S^{2NM-1}$, \textit{i.e.}, $\vert \Psi_i \rangle \sim \mathrm{Uniform}(S^{2NM-1})$. 
Such a sampling scheme for Markov chains is convenient (i) because it can be achieved efficiently in a simple, two-step procedure by drawing $\tilde{\mathbf{x_i}}  \sim \mathcal{N}(\mathbf{0}_{2NM}, \mathbbm{1}_{2NM})$ and then normalizing $\tilde{\mathbf{x_i}}/\left\| \tilde{\mathbf{x_i}}\right\| \rightarrow \mathbf{x_i}$; and (ii) because it does not break the detailed balance condition for Metropolis-Hastings MCMC. 
The proposal distribution $g(\Psi_\mathrm{f} \vert \Psi_\mathrm{i} )$, \textit{i.e.}, the conditional probability of proposing state $\vert \Psi_\mathrm{f} \rangle$ given state $\vert \Psi_\mathrm{i} \rangle$, is constant over the state space and therefore symmetric in its two arguments. 
That is, we accept a new state $\vert \Psi_{i+1} \rangle$ with the Metropolis acceptance criterion
\begin{align}
    A(\Psi_{i+1}\vert \Psi_i) = \min \left( 1, \frac{e^{-\beta'\langle \Psi_{i+1}\vert \hat{H}\vert \Psi_{i+1}\rangle}}{e^{-\beta'\langle \Psi_{i}\vert \hat{H}\vert \Psi_{i}\rangle}}\right).
    \label{eq:acceptance_criterion}
\end{align}
Accepting and rejecting newly proposed system-bath wave functions $\vert \Psi_{i+1}\rangle$ based on Eq.~\ref{eq:acceptance_criterion} according to Algorithm \ref{alg:MCMC} generates Markov chains $\{\Gamma_0, \Gamma_1, ...\}$ that converge in distribution to $P[\Gamma]$. 
In transforming those Markov chains according to $\{\Gamma_i\} \rightarrow \{\mathrm{Tr_B}\{\Gamma_i\}\}$ we obtain samples $\{\sigma_i\}$ that are distributed according to $P[\sigma] = \int[d\Gamma]P[\Gamma]\delta(\sigma - \mathrm{Tr_B}\{\Gamma\})$. 

\begin{algorithm}[H]
\begin{algorithmic}
\caption{Metropolis-Hastings MCMC scheme to sample from $P[\sigma]$.}\label{alg:MCMC}
\State {$\vert \Psi_0 \rangle \sim \mathrm{Uniform}(S^{2NM-1})$ \Comment{Initialize Markov Chain}}
\State{$\sigma_0 \gets \mathrm{Tr_B}\{\vert \Psi_0 \rangle \langle \Psi_0 \vert\}$}
\For{$i < N_\mathrm{samples}$}
\State {$\vert \Psi_{i+1} \rangle \sim \mathrm{Uniform}(S^{2NM-1})$ \Comment{propose new wave function}}
\State {$u \sim \mathrm{Uniform}(0,1)$ \Comment{accept/reject based on Eq.~\ref{eq:acceptance_criterion}}}
\If{$u \leq A(\Psi_{i+1}\vert \Psi_i)$}
    \State {$\vert \Psi_{i+1} \rangle \gets \vert \Psi_{i+1} \rangle$ \Comment{accept}}
\Else
    \State {$\vert \Psi_{i+1} \rangle \gets \vert \Psi_{i} \rangle$ \Comment{reject}}
\EndIf
\State {$\sigma_{i+1} \gets \mathrm{Tr_B}\{\vert \Psi_{i+1} \rangle \langle \Psi_{i+1} \vert\}$}
\EndFor
\State \Return {$\{\sigma_i\}$}
\end{algorithmic}
\end{algorithm}

While this formalism is general, we are here for the quantum circuit composed of a single qubit particularly interested in a sub-coordinate of $\sigma$, namely the $z$-coordinate in the Bloch sphere representation of $\sigma$, $z = \bar{\sigma_z}[\sigma]$. That is, we further transform $\{\sigma_i\} \rightarrow \{\bar{\sigma_z}[\sigma_i]\} = \{z_i\}$, which is distributed according to $P(z) = \int[d\sigma] P[\sigma] \delta(z-\bar{\sigma_z}[\sigma])$. 

\paragraph{Numerical simulation of a qubit weakly coupled to a degenerate spin bath.} 
The qubit circuit, governed by Hamiltonian $\hat{H}_\mathrm{S}=-\frac{\hbar \omega_0}{2}\sigma_z$ in the basis of logical qubit states $\vert 0 \rangle$ and $\vert 1 \rangle$, is immersed in a bath of $D$ (unbiased) qubits with energy $E_0$. That is, the bath Hamiltonian is given by $\hat{H}_\mathrm{B}=E_0 \mathbbm{1}_M$ with $M=2^D$. We assume that the circuit and bath are coupled through $\hat{V}=\lambda \left(\bigotimes_{i=1}^{D+1}\sigma_x \right)$ and that the coupling strength, parameterized by $\lambda$, is very weak (here $\lambda = 10^{-6}\,\mathrm{eV}$). The energy gap for the qubit is set to $\hbar \omega_0 = 0.30\,\mathrm{eV}$ and we fix the temperature of the circuit-bath composite at $k_\mathrm{B}T = 0.45\,\mathrm{eV}$. 

\paragraph{Numerical simulation of a qubit weakly coupled to a thermally truncated harmonic oscillator.} 
The same qubit circuit is now coupled to the displacements of a single harmonic oscillator with energy gap $\hbar \omega = 0.2\cdot \hbar\omega_0$. 
The Hilbert space of the oscillator, $\mathcal{H}_\mathrm{B}$, is thermally truncated according to the Boltzmann distribution at (inverse) temperature $\beta$ to have (finite) dimension, \textit{i.e.}, $\mathrm{dim}(\mathcal{H}_\mathrm{B}) = M$. 
That is, the matrix elements of the $M \times M$-dimensional bath Hamiltonian are given by  $\langle i \vert \hat{H}_\mathrm{B} \vert j \rangle = \hbar \omega (i + \frac{1}{2})\delta_{ij}$ in the basis of oscillator eigenstates $\{\vert i \rangle\}_{i=0}^{M-1}$.
The circuit-bath coupling is mediated by the operator $\hat{V}=\lambda \left( \sigma_x \otimes \hat{x}\right)$, where $\langle i \vert \hat{x} \vert j \rangle \equiv \sqrt{i}\delta_{i,j+1}+ \sqrt{j}\delta_{j, i+1}$ are the matrix elements of the displacement operator $\hat{x}$ of the harmonic oscillator. 
We again set the coupling parameter to $\lambda = 10^{-6}\,\mathrm{eV}$ in order to guarantee the weak-coupling regime.

\begin{acknowledgments}
 H.J.H. and A.P.W. acknowledge the US Department of Energy (DOE), Office of Science, Basic Energy Sciences (BES) for funding this work through award DE-SC0019998.
\end{acknowledgments}

\newpage
\bibliography{LandauerReferences}

\end{document}